\documentclass[revtex4-1]{article}
\usepackage[utf8]{inputenc}
\usepackage[a4paper, total={14cm, 23cm}]{geometry}
\usepackage{graphicx}
\usepackage{amsfonts}
\usepackage{amssymb}
\usepackage{amsmath}
\usepackage{epsfig}
\usepackage{amsthm}
\usepackage{babel}
\usepackage{multicol}
\usepackage{fancyhdr}
\usepackage{slashed}
\usepackage{eucal}
\usepackage{mathrsfs}
\usepackage{mathtools}
\usepackage{authblk}
\usepackage{appendix}
\usepackage{cite}
\usepackage[dvipsnames]{xcolor}
\usepackage{comment}
\usepackage{csquotes}
\usepackage{caption}
\usepackage{subcaption}
\usepackage{hyperref}

\DeclareSymbolFont{bbold}{U}{bbold}{m}{n}
\DeclareSymbolFontAlphabet{\mathbbold}{bbold}

\makeatletter
\DeclareFontEncoding{LS2}{}{\noaccents@}
\makeatother
\DeclareFontSubstitution{LS2}{stix}{m}{n}
\DeclareSymbolFont{xlargesymbols}{LS2}{stixex}{m}{n}
\DeclareMathSymbol{\sumop}{\mathop}{xlargesymbols}{"B3}

\graphicspath{{home/}}

\numberwithin{equation}{section}

\newcommand{\bo}{\boldsymbol}
\newcommand{\smallfrac}[2]{{\textstyle\frac{#1}{#2}}}
\newcommand{\BE}{\begin{equation}}
\newcommand{\EE}{\end{equation}}
\newcommand{\mrm}{\mathrm}
\newcommand{\bb}{\mathbb}

\newcommand{\dd}{\mathrm{d}}

\newcommand{\me}{\mathrm{e}}

\newcommand{\mcal}{\mathcal}

\newcommand{\mn}{\mathnormal}
\newcommand{\del}{\partial}
\newcommand{\eps}{\epsilon}

\newcommand{\mPhi}{\mn\Phi}

\newcommand{\nn}{\nonumber}

\newcommand{\bnab}{\bo{\nabla}}
\newcommand{\im}{\mathrm{Im}}
\newcommand{\re}{\mathrm{Re}}

\newcommand{\ket}[1]{| #1 \rangle}

\newcommand{\bx}{{\bo{x}}}
\newcommand{\by}{{\bo{y}}}
\newcommand{\bxp}{{\bo{x}'}}

\newcommand{\bp}{{\bo{p}}}

\newcommand{\bk}{{\bo{k}}}

\newcommand{\what}{\widehat}

\newcommand{\f}[1]{{\mathscr #1}}

\newcommand{\eqn}[1]{eq. (\ref{#1})}

\makeatletter
\newsavebox{\@brx}
\newcommand{\llangle}[1][]{\savebox{\@brx}{\(\m@th{#1\langle}\)}%
  \mathopen{\copy\@brx\kern-0.5\wd\@brx\usebox{\@brx}}}
\newcommand{\rrangle}[1][]{\savebox{\@brx}{\(\m@th{#1\rangle}\)}%
  \mathclose{\copy\@brx\kern-0.5\wd\@brx\usebox{\@brx}}}
\makeatother

\title{{Simulating Nelsonian Quantum Field Theory}}
\author{Andrea Carosso\thanks{email: acarosso@gwu.edu}}
\affil{\textit{\small{Department of Physics, George Washington University, Washington, D.C., United States}}}
\date{}

\begin{document}

\setlength{\abovedisplayskip}{4pt}
\setlength{\belowdisplayskip}{4pt}

\maketitle

\sloppy

\abstract{We describe the picture of physical processes suggested by Edward Nelson's stochastic mechanics when generalized to quantum field theory regularized on a lattice, after an introductory review of his theory applied to the hydrogen atom. By performing numerical simulations of the relevant stochastic processes, we observe that Nelson's theory provides a means of generating typical field configurations for any given quantum state. In particular, an intuitive picture is given of the field ``beable'' --- to use a phrase of John Stewart Bell --- corresponding to the Fock vacuum, and an explanation is suggested for how particle-like features can be exhibited by excited states. We then argue that the picture looks qualitatively similar when generalized to interacting scalar field theory. Lastly, we compare the Nelsonian framework to various other proposed ontologies for QFT, and remark upon their relative merits in light of the effective field theory paradigm. Links to animations of the corresponding beables are provided throughout.}

\tableofcontents

\newpage

\section{Introduction}

A prominent strategy for supplying standard quantum mechanics (QM) with a clear ontology --- a detailed account of what exists according to the theory\footnote{Bell \cite{Bell-Beables} suggested the particular term ``beable'' for such things (rather than, e.g., ``being'' or ``beer'') to emphasize ``the essentially tentative nature of any physical theory,'' a sentiment we want to reiterate in this work. ``In fact, `beable' is short for `maybe-able'.''} --- is provided by David Bohm's deterministic `hidden' variable theory from 1951 \cite{Bohm-I:1951,Bohm-II:1951}, which constituted a completion of the work of Louis de Broglie from the 1920's (though Bohm arrived at these ideas independently).
For non-relativistic QM it was suggested that, along with the wave function, there exist particles whose trajectories in space are determined by the wave function --- hence the name ``pilot wave theory.'' Although their theory provides a brilliant example of how a detailed picture of quantum processes is possible, there are nevertheless certain unappealing traits of the theory, as usually presented. One such trait is the particle behavior when the wave function has a spatially constant phase, where it is found, for example, that the (spinless) electron in the hydrogen ground state is \emph{motionless} relative to the proton.\footnote{A few caveats: In \cite{Durr:1992}, this problem is avoided by remarking that such eigenstate distributions, in realistic scenarios, arise from complex interactions, during which the wave function is not so simple. Second, once spin is included, the electron is no longer motionless in such states \cite{Colijn:2002,Krekels:2023}.} Another unappealing trait\footnote{Not to everybody. See \cite{Ney} for various perspectives on the status of the wave function. See also \cite{Norsen:2014, Hubert:2018}.} is the apparent need to regard the wave function as a physical entity that exists on a high-dimensional configuration space, rather than the 3-dimensional space of ordinary experience, when many particles are involved. There may be other unappealing traits of Bohm's original theory, but here we shall not treat with them.

% Bohm-I received 5 July 1951
% Fenyes recieved 30 January 1952 ("Eingegangen")
% Weizel received 17 October 1952

Shortly after Bohm's first papers appeared, independent work was being carried out which related the Schr\"odinger equation to diffusion processes \cite{Fenyes, Weizel}, and soon Bohm and Vigier \cite{Bohm-Vigier} joined the discussion by arguing for the possibility of regarding QM as arising from an underlying stochastic process involving particles in interaction with a ``sub-quantum'' medium. A decade later, in 1966, Edward Nelson independently obtained these results and placed them in a rigorous mathematical framework with his ``stochastic mechanics'' \cite{Nelson:1966}. He proposed to \emph{derive} QM by postulating an underlying stochastic process, akin to a Brownian motion, which however is subject to a stochastic generalization of Newton's Second Law.\footnote{Whether Nelson succeeded in deriving QM in its entirety has been argued in the negative by Wallstrom \cite{Wallstrom}, but there is some disagreement \cite{Kuipers-2023}. Proposals to ``complete'' the derivation, assuming the validity of Wallstrom's criticism, have also been made in recent years \cite{Derakhshani:2015,Derakhshani:2019}.} The picture of particle motion as determined by Nelson's theory is that particles bump around in seemingly random ways, similar to how a Brownian particle moves, and the statistical distribution generated by this random motion matches that predicted by standard QM --- the probability density of the electron position converges toward the time-varying quantity $\rho(\bx,t) = |\psi(\bx,t)|^2$. 

Nelson's theory is closely related to Bohm's theory; the equation determining particle motion for Nelson is a stochastic generalization of Bohm's. As a particular consequence, the electron in the ground state of hydrogen is no longer motionless, but bumps about the proton in a random fashion. In their final elaboration of pilot wave theory as described in \emph{The Undivided Universe} \cite{Bohm-Hiley}, Bohm and Hiley themselves admitted that the random motion entailed by the stochastic interpretation is more plausible than the deterministic one:
\begin{quote}
``Such a view of the $s$-state as one of dynamic equilibrium seems to fit in with our physical intuition better than one in which the particle is at rest.'' p. 201.
\end{quote}
In Section 2, the formalism of Nelson's theory will be reviewed, and the results of simulations of Nelson's theory applied to the case of hydrogen will be reported on, displaying explicitly various sample trajectories and demonstrating their consistency with standard QM. For further reviews of stochastic mechanics, see \cite{Guerra:1981,Seiler,Goldstein-Nelson,Guido-Concept}.

In generalizing Bohm's theory to quantum \emph{fields}, there remains a lack of agreement about what the proper ontology should be: whether to take particles, fields, \emph{both}, or something else. And this is reflective of the broader discussion among philosophers and physicists about what the ontological content of QFT could \emph{possibly} be \cite{Baker-2009,Fraser-1,Wallace-Naive,Williams,Egg-Lam,Sebens:2022}. The first significant proposal for QFT was made by Bohm already in 1951 \cite{Bohm-II:1951}, where an ontological interpretation was formulated for bosonic field theories, in which the beables were simply field configurations across space, just as in classical electrodynamics, for example, but which evolved in time under the influence of a wave function\emph{al}. But such a picture does not immediately explain the appearance of particles, or particle-like objects, in the world, since fields are inherently extended objects. Bohm argued that the discrete nature of photons as described by a photon \emph{field}, say, nevertheless emerges in the course of certain experiments \cite{Bohm-II:1951,Bohm-Hiley}.\footnote{See \cite{Bohm-Hiley,Holland}, and \cite{Valentini-Cushing,Dewdney-Cushing,Struyve-Westman} for similar field-based proposals.} Later on, various other proposals were made that regarded the ``primitive ontology'' of QFT as being comprised of \emph{particles} \cite{Bell-Book, Durr:2004, Colin:2007, Nikolic:2009, Deckert:2019,Oldofredi2020}, instead of fields. An appealing feature of these proposals is their explicit incorporation of particle creation and annihilation (or \emph{apparent} creation and annihilation), which is a characteristic property of QFT that ought to be accounted for in any theory hoping to clarify its ontology.

In Section 3, the generalization of Nelson's theory to QFT will be described, and the results of various simulations of the dynamics will be displayed. Although Nelson focused primarily on a stochastic formulation of nonrelativistic QM, the field-theoretical generalization of his theory was explored finally in the 1970's and 1980's \cite{Guerra:1973,Guerra:1981}. Since then, however, the subject has seen little activity.\footnote{What seems to have stuck around for longer was the related enterprise of Stochastic Quantization (SQ) -- but this too has mostly fallen out of fashion. An important difference between SQ and Nelson should be stressed: the ``time'' of SQ is a fictitious ``simulation time,'' whereas the time for Nelson is a real, physical time. See \cite{Koide:2013} for a recent perspective on Nelson-Yasue QFT; and see \cite{Damgaard} for an introduction to SQ.} The beables of this theory are fields, similar to Bohm's original proposal, except that the evolution with time of the field is stochastic rather than deterministic. We will discuss what the vacuum state of a free-field QFT is like according to this theory, and we will further observe that the field beable corresponding to certain $n$-particle states in free QFT indeed have particle-like characteristics, namely, spike-like (or lump-like) regularities at the locations of the ``particles.'' Just as the stochasticity of the non-relativistic theory yielded a dynamical description of electron motion in stationary states, the particle-like structures present in certain time-independent $n$-particle QFT states are dynamically generated in the stochastic theory; Bohm's version, applied to the same states, yields field beables that need not have these particle-like characteristics (for a single field configuration). A genuinely time-dependent state will also be examined, and the corresponding Nelsonian and Bohmian beables will be compared. We will argue that qualitatively similar properties continue to hold even for interacting scalar field theory, and we will comment on the account of particle creation as well as the extension to fermionic field theory.

The Nelsonian account of QFT has a further appealing feature. In the modern practice of QFT, the philosophy of effective field theory (EFT) has taken hold, beginning in 1970's with the advent of Wilsonian RG \cite{Wilson:1973}. The philosophy to emerge regards QFT, in nearly every real-world application, as being a merely effective theory, in the sense that it is expected to be valid only on a limited domain of distance and energy scales \cite{Lepage:1989,Srednicki,Cao-Schweber}. In particular, this modern understanding regards each theory as having some ultraviolet (UV) energy scale beyond which it ceases to remain a good description of physical processes. Thus, for example, Fermi's theory of weak interactions is an effective theory of leptonic processes valid at scales lower than the W and Z boson masses; chiral perturbation theory is an effective theory of pion-nucleon interactions at scales below the proton mass; and it is widely believed that the Standard Model itself is the effective theory.
Now, since Nelson's theory regards QM as an effective theory to begin with, as we will describe in Section 2, one might hope that it will provide a more appropriate framework for the interpretation of modern QFTs.

% ====================================================================

% =================================================================

\section{Nelson's Stochastic Mechanics}

Here we briefly review the formulation of Nelson's stochastic mechanics for a single particle, and how it was later interpreted by Bohm and Hiley. A description of the hydrogen atom according to this theory is then provided, together with results from the simulation of a discretized version thereof.

\subsection{Mathematical Formulation}

Nelson proposed that for any wave function $\psi(\bo x,t)$ evolving according to Schr\"odinger's equation (in the position space representation) there exists a stochastic process which generates the same statistical properties entailed by $\psi(\bo x,t)$, i.e. whose probability density $P(\bx,t)$ for particle positions matches $\rho(\bx,t)=|\psi(\bo x, t)|^2$, and which satisfies the continuity equation
\BE
\del_t \rho(\bx,t) + \bnab \cdot \bo j(\bx,t) = 0,
\EE
$\bo j(\bx,t)$ being the probability current density associaed with $\psi(\bx,t)$.
The stochastic process, in particular, is defined by a Langevin equation of the form\footnote{Although the derivative $\dot{\bx} = \dd \bx /\dd t$ here is ill-defined, rigorous definitions exist via Weiner processes \cite{Pavliotis:2014,Damgaard}, which are here swept aside to ease the presentation.}
\BE \label{langevin}
\dot{\bo x}(t) = \bo b(\bo x, t) + \bo \eta(t),
\EE
where $\bo x = \bo x(t)$ is the instantaneous location of a particle, and $\bo b(\bx,t)$ is the \emph{drift} vector field, to be specified shortly. $\bo \eta_t$ is Gaussian random noise, which is defined by its first two moments:
\begin{align}
\mrm{E}[\eta^i(t)] = 0, \quad \mrm{E}[\eta^i(t) \eta^j(s)] = 2 \nu \delta^{ij} \delta(t-s),
\end{align}
where $\nu = \hbar/2m$ is the diffusion coefficient, and E denotes expectation value with respect to the Gaussian noise measure.
Guided by an analogy with the classical Ornstein-Uhlenbeck theory, Nelson subjected the particle to a further condition by demanding that a certain ``mean acceleration'' of the particle should satisfy Newton's Second Law:
\BE
m \overline {\bo a}(t) = - \bo\nabla V(\bo x,t),
\EE
where $V$ is an external potential. He chose $\overline {\bo a}(t) = (D D_* + D_* D) \bo x(t)$, which constitutes a time-reversal symmetric definition, where $D,D_*$ are the so-called ``mean forward'' and ``mean backward'' differential operators associated with the process \eqn{langevin}.\footnote{Here $(Df)(\bx,t):= \lim_{\eps\to 0^+} \eps^{-1}\mrm{E}[f(\bx_{t+\eps},t+\eps)-f(\bx_t,t)|\bx_t=\bx]$. One can define a ``backward'' stochastic process associated with the forward process, i.e., a rule relating the current positions to prior positions in a probabilistic fashion, and an associated backward derivative $D_*$; see \cite{Nelson:1966,Guerra:1981,Bohm-Hiley-Stoch}.} From these ingredients, Nelson argued that the drift can be written in terms of two functions $R$ and $S$ as
\BE \label{nelson-drift}
\bo b(\bo x, t) = \frac{1}{m} \Big( \bo \nabla R(\bo x, t) + \bo \nabla S(\bo x, t)\Big),
\EE
where $R$ and $S$ satisfy (letting $\f R =\exp(R/\hbar)$)
\BE
\del_t S + \frac{1}{2m} |\bnab S|^2 + V(\bx,t) - \frac{\hbar^2}{2m} \frac{\Delta \f R}{\f R} = 0,
\EE
\BE
\del_t \f R^2 + \frac{1}{m} \bnab \cdot (\f R^2 \bnab S) = 0.
\EE
Since these equations are the real and imaginary parts of the Schr\"odinger equation (for a spinless particle in a potential $V(\bx,t)$) under the polar decomposition $\psi = \exp[(R + i S)/\hbar]$, Nelson claimed to have derived QM (see footnote 4 for a qualification, however). Alternatively, the drift above can be written directly in terms of the wave function:
\BE \label{nelson-drift-2}
\bo b(\bo x,t) = \frac{\hbar}{m} \big( \mrm{Re} + \mrm{Im} \big) \frac{\bnab \psi(\bo x,t)}{\psi(\bo x,t)}.
\EE
The first term in eqs. (\ref{nelson-drift}, \ref{nelson-drift-2}) is called the ``osmotic velocity,'' since it plays a role similar to the osmotic term in typical diffusion processes; it tends to steer particles toward maxima of $\exp(R(\bo x,t)/\hbar)$, the amplitude of the wave function. The second term is the same as in the de Broglie-Bohm theory guidance law, $m \dot{\bx} = \bnab S$; it tends to guide particles perpendicular to the wave fronts of $\exp(iS(\bo x,t)/\hbar)$.

The trajectories determined by stochastic equations such as eq. (\ref{langevin}) are continuous but not differentiable: the position is well-defined, but the instantaneous velocity is not, except in an average sense, $\overline{\bo v} := D \bo x = \bo b$. The case $\bo b=0$ is just a standard Brownian motion as proposed by Einstein in 1905, although the material origin of the fluctuations in the velocity are no longer the molecules in a solution, but arise from interactions with some underlying background field, or a bath of minute particles, the detailed properties of which are left unspecified. In this sense, we may regard Nelson's theory as an \emph{effective} theory, valid on time/distance scales large enough that the effect of the background field is captured by eq. (\ref{langevin}), in which case standard QM itself would be an effective theory. As such, the velocity may ultimately be well-defined on short-enough time scales, but not on the course-grained timescale.

The limit of zero-noise and zero-osmotic term yields the de Broglie-Bohm guidance law, $\dot{\bx} = \frac{1}{m}\bo \nabla S(\bo x, t)$, which formed the basis of Bohm's \emph{deterministic} hidden variable theory: given an initial condition $\bx(0)$, the future positions $\bx(t)$ are entirely determined, and probabilities enter as a consequence of uncertainty in the initial condition. With nonzero noise, however, the motion is no longer deterministic; the stochastic process generates a probability distribution $P(\bo x,t)$ of particle positions that approaches the density $\rho(\bo x,t) = |\psi(\bx,t)|^2$ for times much larger than some characteristic relaxation time $\tau_\mrm{eq}$, unique to each process. Since this limit is nonetheless time-varying, in general, and since the distribution matches that implied by the wave function, one calls this ``quantum equilibrium,'' rather than just ``equilibrium.'' Throughout this work we will refer to the above Langevin equation as the ``stochastic guiding law.''

Bohm and Hiley \cite{Bohm-Hiley-Stoch} criticized Nelson's theory for assuming a particular definition of the acceleration, which they felt was ad hoc, since other definitions are possible which nevertheless result in Schr\"odinger's equation, and which obey the time-reversal symmetry desired by Nelson.\footnote{Yasue \cite{Yasue} has shown that Nelson's Second Law follows from a stochastic variational principle, together with the demand of positive semidefinite mean kinetic energy.} Instead, they reinterpreted the theory, much in line with standard Bohmian mechanics, to merely \emph{postulate} the Schr\"odinger equation, and provide a definite ontology by asserting the existence of a particle undergoing the random motion determined by the wave function via the stochastic guiding law. In so doing they drop the postulate of a stochastic Newton's Second Law, and furthermore avoid the Wallstrom criticism (see footnote 4). In effect, they take a less ambitious perspective than Nelson did. In what follows, we adhere to the interpretation of Bohm and Hiley as a matter of simplicity, but continue to refer to it as Nelson's stochastic mechanics.

\subsection{Discrete Formulation}

Throughout this work, we will supplement the discussion by displaying results of numerical simulations of Nelson's theory. We therefore describe in brief how to suitably formulate Nelson's dynamics for simulation. See also \cite{McClendon:1988,Nitta:2008,Hardel:2023} for previous simulations of stochastic mechanics.

We discretize the time variable into steps of equal duration $\eps$, and define a dimensionless Gaussian noise $\what{\bo \eta}_t$ with mean zero and variance 1, related to the previously defined noise by sending $\bo \eta(t) \to \sqrt{2\nu/\eps} \; \what{\bo \eta}_t$. The discrete Euler step for the Langevin equation becomes, for all $t = n\eps$ with $n\geq0$ an integer,
\BE
\bo x_{t+\eps} = \bo x _t + \eps \bo b(\bo x_t, t) + \sqrt{2 \nu \eps} \; \what{\bo \eta}_t.
\EE
Here the factor $2\nu = \hbar/m$ explicitly appears in the noise term. Since the time step $\eps$ is finite, the actual (quantum-) equilibrium distribution of the discrete process will deviate from the desired distribution by terms of order $\eps$, and therefore one is ultimately concerned with taking the limit $\eps \to 0$, or simulating instead with an ``improved'' discrete equation that eliminates leading effects, such as the stochastic Runge-Kutta algorithm in \cite{Drummond:1983} or \cite{Batrouni}. In this work we restrict ourselves to the simple Euler discretization, and merely make note of the behavior of repeated simulations with decreasing $\eps$.

When implementing such simulations, it is most convenient to work with dimensionless positions and times; in any given problem, there will be a natural length scale $(a_0)$ and time scale $(\tau_0)$, so the quantities entering the simulation will be given in these units. In the case of hydrogen, for example, the natural unit for distance is the Bohr radius $a_0$, and natural times are in units of the ``Bohr time'' $\tau_0 = m a_0^2/\hbar$.

\subsection{Application to Hydrogen}

\begin{figure}
\centering
\begin{subfigure}{.5\textwidth}
  \centering
  \includegraphics[width=\linewidth]{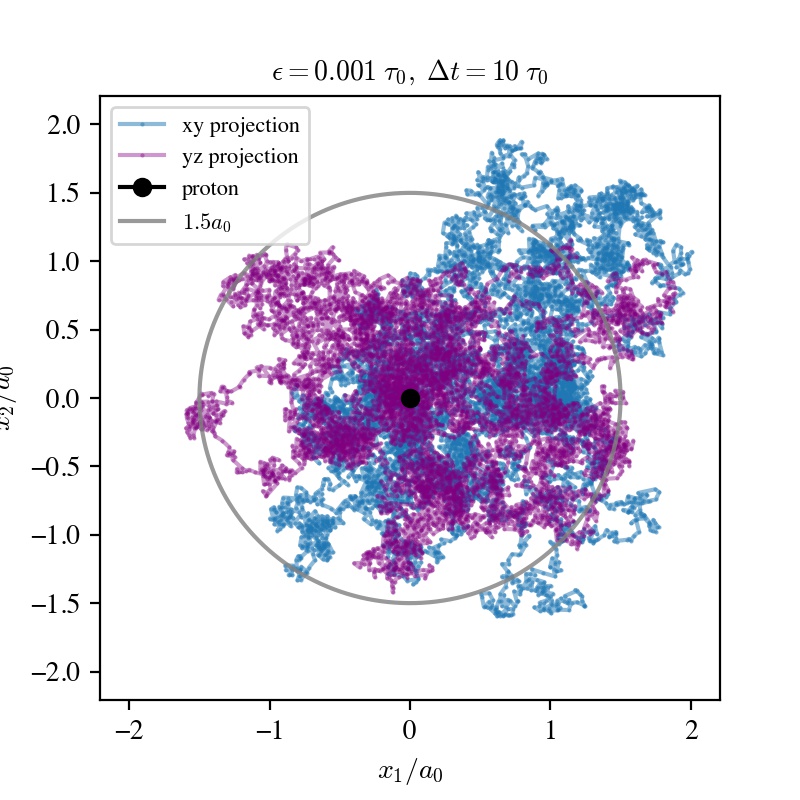}
  \caption{$\psi_{100}(\bx,t)$}
  %\label{fig:sub1}
\end{subfigure}%
\begin{subfigure}{.5\textwidth}
  \centering
  \includegraphics[width=\linewidth]{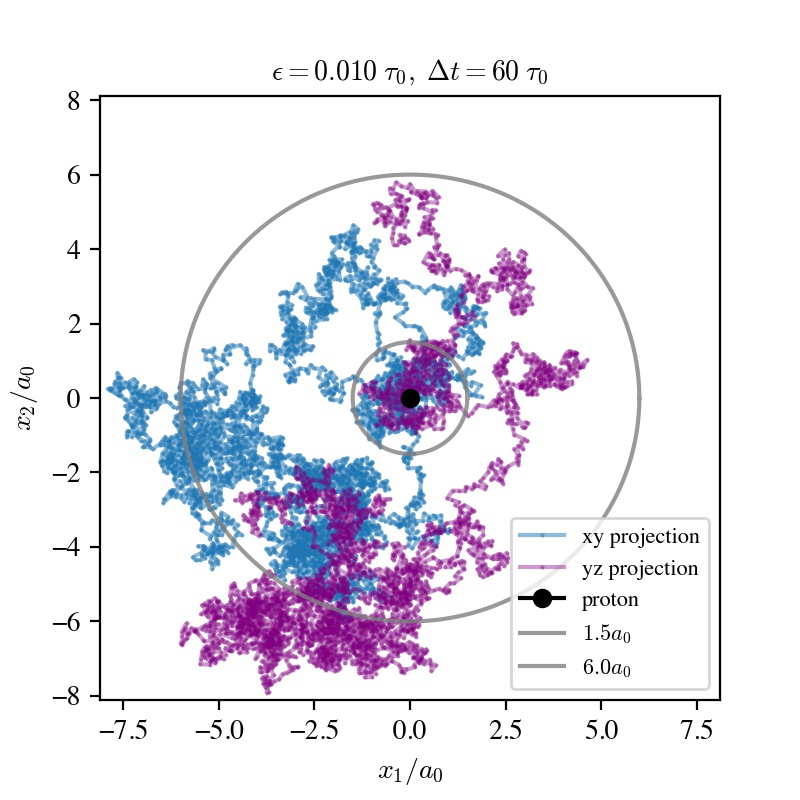}
  \caption{$a\psi_{100}(\bx,t) + b \psi_{200}(\bx,t)$}
  %\label{fig:sub2}
\end{subfigure}
\caption{\small{Projections to the x-y and y-z planes of sample trajectories of an electron in the hydrogen ground state (left), and a superposition of ground and first excited states (right). The time steps and total plotted time are noted above each figure. The average radius for $\psi_{100}$ alone is $1.5 a_0$, and that of $\psi_{200}$ alone is $6 a_0$.}}
\label{fig:trajectories}
\end{figure}

We now consider the behavior of a (spinless) ``electron'' in the hydrogen atom according to Nelson's theory. First consider the ground state, whose wave function has the well-known form
\BE
\psi_{100}(\bo x, t) = \frac{1}{\sqrt{\pi a_0^3}} \; \me^{-r/a_0 - i E_1 t/\hbar},
\EE
where $r = \| \bo x\|$ is the distance to the proton and $E_1 = - me^4/2(4\pi\eps_0)^2 \hbar^2$. In standard Bohmian mechanics, the position-independence of the imaginary part of the phase, in this state, yields a motionless particle: $m\dot{\bo x} = \bo\nabla S = 0$. However, in Nelson's theory, there is a contribution from the osmotic velocity and the random background noise, yielding a stochastic guiding law
\BE
\dot{\bo x} = - \frac{\hbar}{m a_0} \frac{\bo x}{r} + \bo \eta_t.
\EE
We see that the displacement over any short span of time tends toward the origin, but the additional random kicks given to the particle by the background noise prevent it from settling there. By dividing both sides by $a_0$, we obtain natural units for the simulation time, $\tau_0 = m a_0^2/\hbar \approx 2.42 \times 10^{-17}$ sec, which we might call the ``Bohr time.'' In Figure \ref{fig:trajectories} (a), we display a sample trajectory of the electron.\footnote{See \cite{vid-1} for an animation of this motion.} The average radius $\overline{r(t)}$ of this motion at each time, determined from an ensemble of 100 such trajectories\footnote{Or, from an average along a single trajectory, since this state has a time-independent density. In this case one must properly deal with autocorrelations in the time series in order to obtain correct stochastic error estimates; see, e.g., \cite{Montvay}} is displayed in Figure 2 (a), showing consistency with the expected value $1.5 a_0$.

Next we consider two states that have more complicated wave functions: one with nodes, and another which is a superposition of states; we will encounter qualitatively similar phenomena in the generalization to field theory in Section 3. Consider the first excited state of the system, with $\ell = 0$ angular momentum,
\BE
\psi_{200}(\bo x, t) = \frac{1}{\sqrt{32 \pi a_0^3}} \Big(2 - \frac{r}{a_0} \Big) \; \me^{-r/2a_0 - i E_2 t/\hbar}.
\EE
Notice that the state now has a node: $\psi_{200}$ vanishes on the sphere $r=2a_0$. Moreover, the phase is again spatially constant, so $\bo\nabla S=0$. The stochastic process is determined solely by an osmotic term and noise; the corresponding stochastic guiding law is
\BE
\dot{\bo x} = - \frac{\hbar}{m a_0} \cdot \frac{\bo x}{2r} \cdot \frac{4-r/a_0}{2-r/a_0} + \bo \eta_t.
\EE
The node of the wave function results in a singularity in the drift at $r=2a_0$. Physically, this means that if, at some moment in time, the particle finds itself close to that radius, the drift it feels will be large, and directed \emph{away} from the node: it is repelled away from this region of space. In discretized form, the node leads to an instability in the numerical integration: since the time step is finite, it is possible for the particle to accidentally get displaced arbitrarily close to the node, resulting in a massive jump in the particle location; it can then take many timesteps for the electron to arrive back in the vicinity of the proton, but such excursions make a reliable estimation of statistical properties difficult. However, we can, for example, empirically count the average number of anomalous jumps per unit time as a function of stepsize $\eps$; we find a rapidly decaying behavior as $\eps \to 0$, indicating that the anomalies become rare in the time-continuum limit, as suggested by Nelson in \cite{Nelson-Book}. See Figure \ref{fig:100anomalies}. 

\begin{figure}
\centering
\begin{subfigure}{.5\textwidth}
  \centering
  \includegraphics[width=1.1\linewidth]{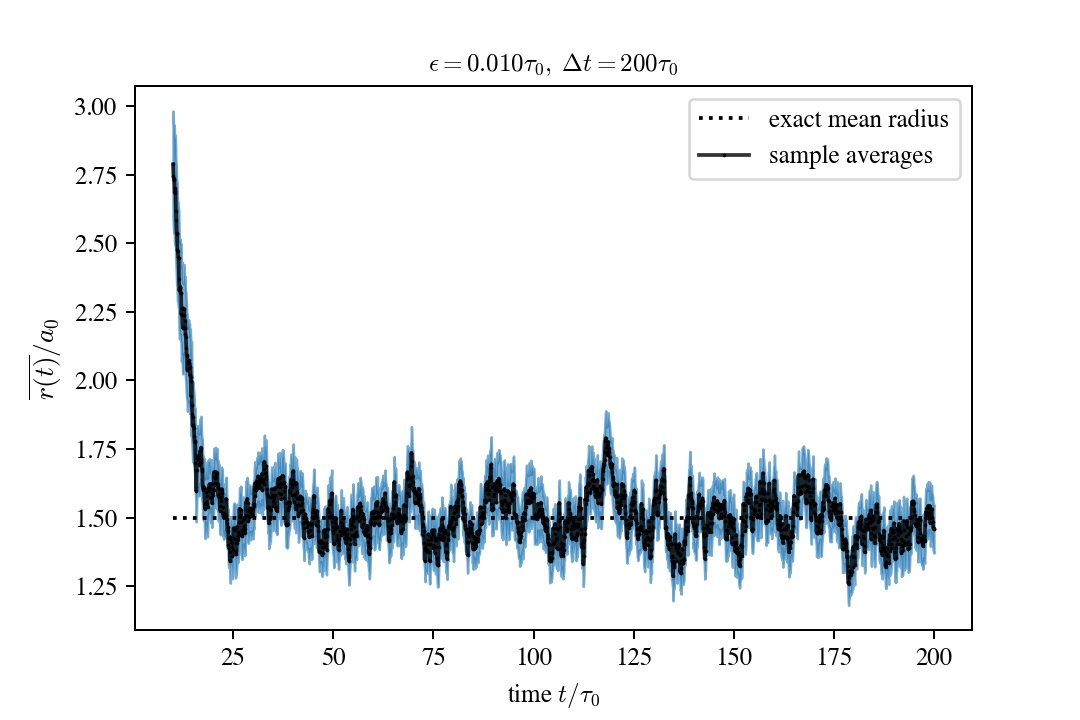}
  \caption{$\psi_{100}(\bx,t)$}
  %\label{fig:sub1}
\end{subfigure}%
\begin{subfigure}{.5\textwidth}
  \centering
  \includegraphics[width=1.1\linewidth]{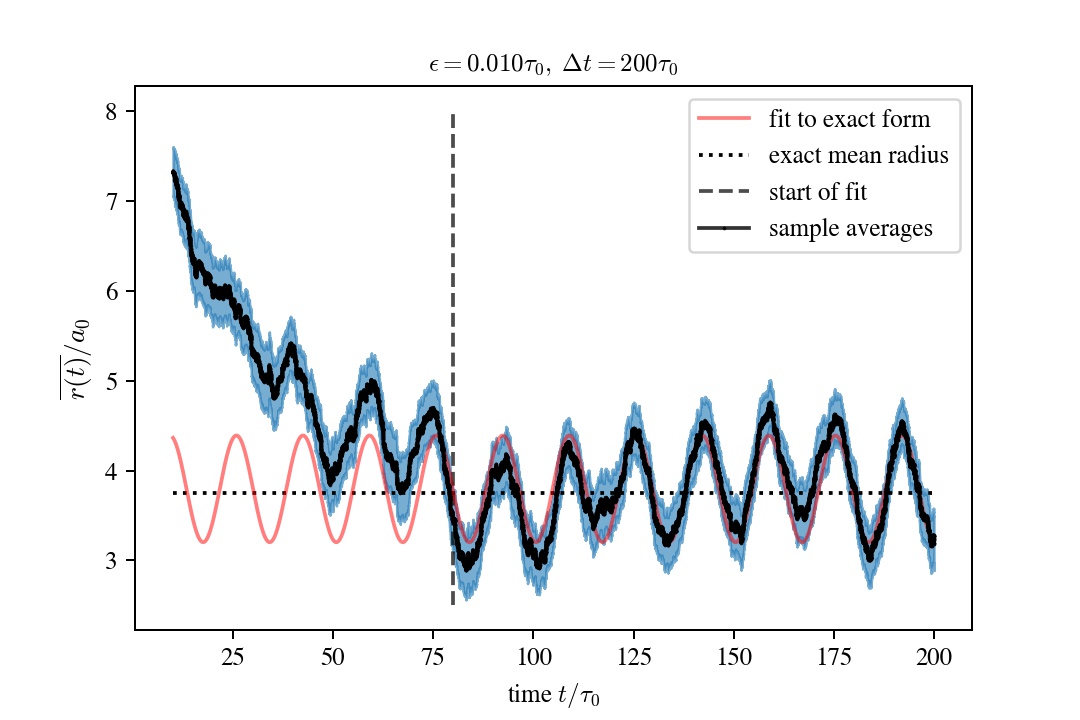}
  \caption{$a\psi_{100}(\bx,t) + b \psi_{200}(\bx,t)$}
  %\label{fig:sub2}
\end{subfigure}
\caption{\small{Averages over 100 sample trajectories of the distance of the electron from the proton in the hydrogen ground state (left), and the same superposition of ground and first excited states as Fig. 1 (b) (right). The initial electron positions were taken to have radius $10 a_0$ in each case; the approach to quantum equilibrium is evident; in the case of (b) this limit is time-dependent. Blue bands are the standard error on the mean radius at each time. With increasing number of samples, the errors decrease.}}
\label{fig:radii}
\end{figure}

\begin{figure}
    \centering
    \includegraphics[scale=0.55]{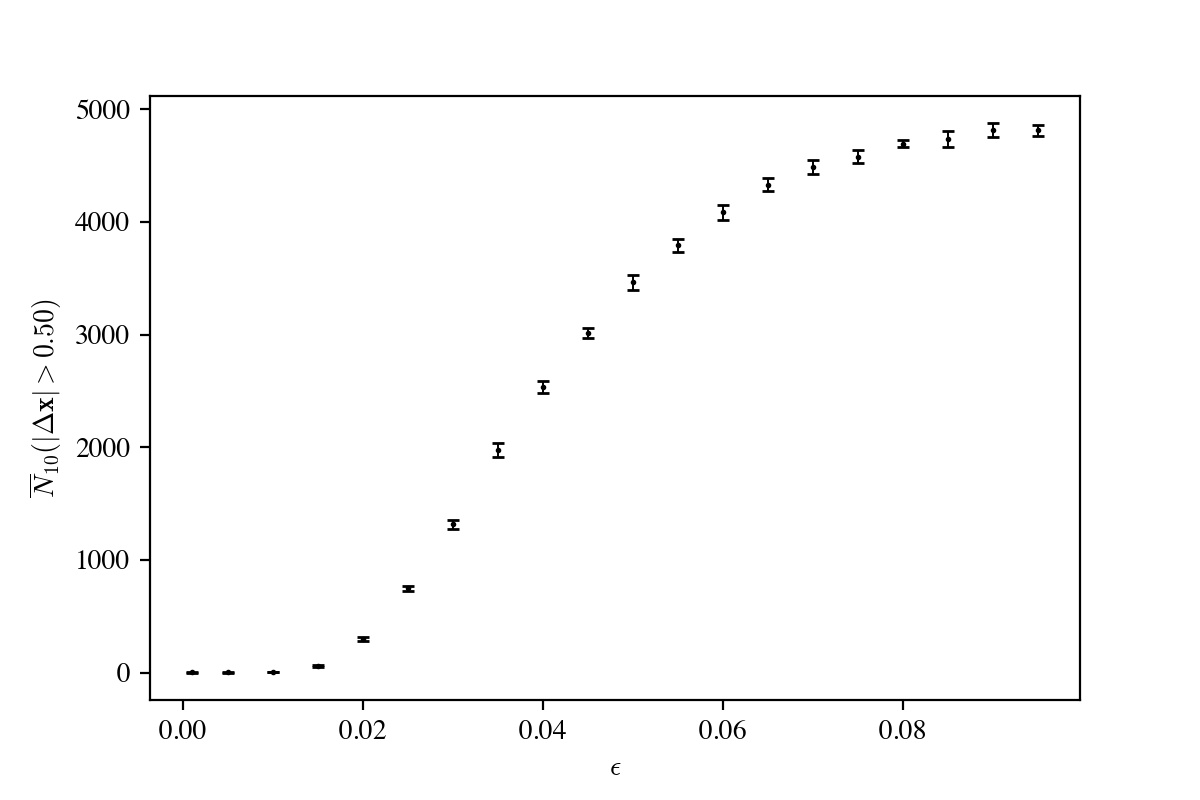}
    \caption{\small{Average number of anomalous jumps $\overline N$, here defined as changes $\|\bx_{t+\eps}-\bx_t\| > 0.5 a_0$, during simulations of electron motion in the state $\psi_{200}$. The physical time was kept fixed at $\Delta t = 1000 \tau_0$ for each timestep $\eps$. The averages were taken over 10 sample trajectories for each $\eps$, and uncertainty bars are the standard deviation.}}
    \label{fig:100anomalies}
\end{figure}

The last example we discuss is a time-dependent superposition of the two states considered above, with arbitrary coefficients:
\BE \label{superposition}
\psi(\bo x,t) = a \; \psi_{100}(\bo x,t) + b \; \psi_{200}(\bo x,t),
\EE
where $|a|^2 + |b|^2=1$. This state now has a nontrivial phase field $S(\bo x,t)$. A sample trajectory of the electron in such a state is provided in Figure 1 (b).
Since the state is time-dependent, one must compute ensemble averages by averaging over different sample trajectories at each time of interest. In Figure \ref{fig:radii} (b) we show the average radius as a function of time, computed from an average over 100 sample trajectories, in the case of $a=b=1/\sqrt{2}$. The exact behavior is a standard interference function,
\BE
\langle r \rangle_{\psi_t} = \int \dd^3 x \; r |\psi(\bo x,t)|^2 = a_0 \Big( \smallfrac{3}{2} a^2 + 6 b^2 + 2ab A \cos\big((E_2-E_1) t/\hbar\big) \Big),
\EE
where $A$ is a constant that may be worked out.
We find good fits to this form ($\chi^2/\mrm{dof} \approx 1$), so long as we shift the argument of the cosine to allow for a nonzero relaxation time to quantum equilibrium. We emphasize again that this ``equilibrium limit'' itself is a time-varying probability distribution, and the simulation time is a ``real time,'' as opposed to the ``imaginary time'' commonly used in lattice simulations.\footnote{Hence there is no ``sign problem,'' as it is known in the lattice community \cite{Alexandru:2020}, for Schr\"odinger picture expectation values. However, whether Nelson's theory can be used to study problems whose wave functions are \emph{not} known beforehand is an open question, and there is a debate about the ability to compute Heisenberg picture expectation values, like $\langle \what x(t) \what x(0) \rangle$, in Nelson's theory. See \cite{Blanchard:1986,Derakhshani:2022} for the debate.}

\section{Nelsonian Quantum Field Theory}

The stochastic mechanics of a single particle, outlined above, can be readily generalized to the many-particle case \cite{Nelson:1966, Guerra:1973, Bohm-Hiley-Stoch}. Mathematically, a lattice-regularized quantum field theory has the same structure as a system of interacting particles; the simplest case of a free field theory can, in this way, be viewed as a system of coupled harmonic oscillators. In the field theory, the different oscillators are interpreted as field values at definite locations in position (or momentum) space. As such, one can formulate a Nelsonian theory of quantum fields on a lattice, borrowing in fact much of the formalism of traditional lattice QFT \cite{Montvay}. Since this formulation is, however, carried out in the real-time Hamiltonian framework (as opposed to the Euclidean path integral framework), and in particular, the Schr\"odinger representation of the field theory, we first give an outline of the ingredients of that formalism and make note of our particular conventions in doing so. Readers familiar with the Schr\"odinger picture may skip to the next subsection.\footnote{We refer the reader to \cite{Hatfield,Jackiw,Holland} as standard references on the Schr\"odinger picture.}

\subsection{The Schr\"odinger Picture} One begins with the standard canonical quantization of a field $\phi_\bx$ on a lattice, which we take to be a real scalar out of simplicity. The classical field, together with its conjugate field-momentum $\pi_\bx$, are promoted to operators satisfying the canonical commutation relation,
\BE
[\what \phi_\bx, \what \pi_\by] = i a^{-d} \delta_{\bx,\by},
\EE
where $\bx, \by$ are lattice sites in $d$ spatial dimensions, $a$ is the lattice spacing, and $\delta$ is the Kronecker delta. We have set $\hbar=c=1$.
Commutators of $\what \phi$'s among themselves vanish, and likewise for the momenta $\what \pi$. We further assume that the lattice is a periodic box with finite extent $N$ in every direction, so the physical volume is $V = L^d = (aN)^d$. Field operators in momentum space are defined by the lattice Fourier transform,
\BE
\what \phi_\bx = \frac{1}{V} \sumop_\bp \me^{i\bp \bx} \what \phi_\bp, \quad  \what \pi_\bx = \frac{1}{V} \sumop_\bp \me^{i\bp \bx} \what \pi_\bp,
\EE
and we note that
\BE
a^d \sumop_\bx \me^{i\bp \bx} = V \delta_{\bp,0}, \qquad \frac{1}{V} \sumop_\bp \me^{i\bp \bx} = \frac{1}{a^d} \delta_{\bx,0}.
\EE

The (field space) Schr\"odinger representation is obtained by realizing the algebra above by the operations
\BE
\what \phi_\bx = \phi_\bx, \quad \what \pi_\bx = -i a^{-d} \frac{\del}{\del \phi_\bx} \; ,
\EE
acting on wave \emph{functionals} $\Psi(\phi) = \langle \phi | \Psi \rangle$.\footnote{Strictly speaking, these are just ordinary functions of many variables $\phi_\bx$, but we use the term ``functional'' by convention due to the similarity with continuum formulations of QFT \cite{Hatfield}.} This realization of the algebra may be contrasted with its ``field-momentum space'' counterpart, where $\what \pi$ acts by scalar multiplication and $\what \phi$ acts by differentiation, the relation to the field space formulation being provided by the Fourier transform. We will shortly show how the Fock space algebra is realized in the field space as well. The Hilbert space is determined by the inner product of wave functionals,
\BE
\langle \Phi| \Psi \rangle = \int [\dd \phi] \; \Phi(\phi)^* \Psi(\phi), \quad [\dd \phi] = \prod_\bx \dd \phi_\bx,
\EE
which is unproblematic on a lattice.

The classical dynamical system is determined by specifying a  Hamiltonian functional on phase space,
\BE
H(\phi,\pi) = a^d \sumop_\bx \Big[ \; \frac{1}{2} \pi_\bx^2 + \frac{1}{2} \sumop_{k=1}^d (\hat \del_k \phi_\bx)^2 + \frac{m^2}{2} \phi_\bx^2 + \mcal V(\phi_\bx) \; \Big],
\EE
which is just a lattice discretization of the traditional Hamiltonian for continuum free fields. $\pi_\bx = \del_t \phi_\bx$ is the classical canonical field momentum obtained from a Lagrangian.
$H$ is then quantized to obtain the Hamiltonian operator $\what H$ by replacing $\pi \to \what \pi$ and $\phi \to \what \phi$. Above we introduced the finite difference operator $\hat \del$ defined by $\hat \del_k \phi_\bx = (\phi_{\bx + a\bo\me_k} - \phi_\bx)/a$, and $\mcal V(\phi)$ is the potential which defines the type of interaction we are considering; for example, for free fields, $\mcal V=0$, while for quartic interactions, $\mcal  V(\phi_\bx) = \lambda \phi_\bx^4/4!$. The quantum Hamiltonian then reads
\BE \label{ham}
\what H = a^d \sumop_\bx \Big[ - \frac{1}{2 a^{2d}} \frac{\del^2}{\del \phi_\bx^2} + \frac{1}{2} \sumop_{k=1}^d (\hat \del_k \phi_\bx)^2 + \frac{m^2}{2} \phi_\bx^2 + \mcal V(\phi_\bx) \Big].
\EE
The dynamical evolution of wave functionals is then determined by the Schr\"odinger equation,
\BE \label{Schro}
i \del_t \Psi(\phi;t) = \what H \Psi(\phi;t),
\EE
given some initial condition $\Psi(\phi;0) = \Phi(\phi)$. We remark that the standard (lattice) path integral formalism is obtained from this one by Trotterizing the time evolution operator, $U_t = \exp(-i \what H t)$, and inserting completeness relations in field space, as described in many textbooks \cite{Hatfield,Montvay}, and is therefore intimately related to the Schr\"odinger representation.

In the case of free field theory, the eigenstates of the Hamiltonian can be obtained by introduction of the creation and annihilation operators $\what a_\bp, \what a_\bp^\dag$, which obey the algebra
\BE
[\what a_\bp, \what a_\bk^\dag] = 2 \omega_\bp V \delta_{\bp,\bk},
\EE
where $\omega_\bp = \sqrt{\hat \bp^2 + m^2}$ with $\hat p_k = (2/a) \sin (p_ka/2)$, and we have adopted the relativistic normalization convention  \cite{Srednicki}. In terms of these operators, the field operator, for example, is given by
\BE
\what \phi_\bx = \frac{1}{V} \sumop_\bp \frac{1}{2\omega_\bp} \big[ \what a_\bp \me^{i\bp\bx} + \what a_\bp^\dag \me^{-i\bp\bx} \big],
\EE
with a similar formula for $\what \pi_\bx$. The free-field Hamiltonian ($\mcal V=0$ in eq. \ref{ham}) then reads
\BE
\what H_0 = \frac{1}{V} \sumop_\bp \what a^\dag_\bp \what a_\bp + E_0,
\EE
with $E_0=\frac{1}{2}\sumop_\bp \omega_\bp$, and the eigenstates are of the form
\BE
|n;\bp_1,\dots,\bp_n \rangle = C_{n\bp} \; \what a_{\bp_1}^\dag \cdots \what a_{\bp_n}^\dag |0\rangle,
\EE
having energies $E_n(\bp_1,\dots,\bp_n) = E_0 + \sumop_{i} \omega(\bp_i)$. In the Schr\"odinger representation, these states are functionals in $\phi_\bp$. The vacuum state is a Gaussian that can be exactly determined:
\BE \label{free-vacuum}
\Psi_0(\phi) = \langle \phi | 0 \rangle = \Big[\prod_\bp \frac{\omega_\bp}{\pi} \Big]^{\frac{1}{2}} \exp\Big[ -\frac{1}{2V} \sumop_{\bp} \phi_\bp \omega_{\bp} \phi_{-\bp} \Big].
\EE
Excited states with $n>0$ are obtained by applying creation operators to the vacuum, each of which is a differential operator,
\BE
\what a_{\bk}^\dag = \omega_\bk \phi_{-\bk} - V \frac{\del}{\del \phi_\bk}.
\EE
For example, the 1- and 2-particle states are, up to normalization,
\begin{align}
\Psi_1(\phi;\bk) &= \what a^\dag_\bk \Psi_0(\phi) \nn \\
& = 2 \omega_\bk \phi_{-\bk} \Psi_0(\phi) \\
\Psi_2(\phi;\bk_1,\bk_2) & =  \what a^\dag_{\bk_1} \what a^\dag_{\bk_2} \Psi_0(\phi), \nn\\
& =  \big[ 2\omega_{\bk_1} 2\omega_{\bk_2} \phi_{-\bk_1}\phi_{-\bk_2} - 2\omega_{\bk_1} V \delta_{\bk_1,-\bk_2} \big] \Psi_0(\phi),
\end{align}
where $\Psi_n(\phi;\bp_1,\dots,\bp_n) = \langle \phi|n;\bp_1,\dots,\bp_n \rangle$ in Dirac notation.\footnote{One may either compute the norms abstractly using the operator algebra, or by evaluating the functional integral directly, for which standard Gaussian integration formulas apply.} The free-field evolution with time of any of these states can be simply determined by using the property $U_t^\dag a_\bp U_t = \me^{-i\omega_\bp t} a_\bp$.

Conventional QFT often speaks in terms of a particle interpretation, especially in its heavy emphasis on perturbation theory about the free-field limit, and in the use of asymptotic free particle states. The weight behind this interpretation lies in the particle-like properties of the Fock space states $\ket{n,\{\bp_i\}}$, which involve a discrete number of ``quanta'' and have an associated momentum for each quantum. This picture is not totally satisfying, however. First, by associating a particle ontology with the Fock states, it is not clear whether there is any role to be played by physical \emph{fields}. Second, once a concrete realization of the algebra is given via the Schr\"odinger picture, these states turn out to be wave functionals, which do not seem to be very particle-like, being as they are functions of \emph{field configurations}. Lastly, it is unclear what is going on at \emph{intermediate} times, in the non-asymptotic region where there is expected to be significant interaction. In the next sections, we will see how the Nelsonian account of QFT fills in some of these details, and we will mention the corresponding Bohmian pictures.

\subsection{Nelsonian QFT}

Borrowing the mathematical equivalence of a lattice field theory with a many-particle quantum system, we can immediately write down the stochastic process which determines the evolution of the field beable \cite{Guerra:1973,Guerra:1981}. Temporarily restoring $\hbar$ and $c$, we make the replacement $1/m \to c^2/a^d$ and $\del/\del x^i \to \del/\del \phi_\bx$ in the equations of Section 2, since the kinetic energy operator is $-(\hbar^2 c^2/2a^d) \sumop_\bx \del^2/\del \phi_\bx^2$. The Langevin equation in the field theory is then
\BE \label{field-langevin}
\del_t \phi_\bx(t) = b_\bx^\Psi(\phi;t) + \eta_{\bx}(t),
\EE
where the noise is defined by
\BE
\mrm{E}[\eta_\bx(t)] = 0, \quad \mrm{E}[\eta_{\bx}(t) \eta_{\by}(s)] = 2 \nu a^{-d}\delta_{\bx,\by} \delta(t-s),
\EE
with $\nu = \hbar c^2/2$,\footnote{The diffusion coefficient $\nu$ in Nelsonian QFT is therefore independent of any particle mass, or the Klein-Gordon mass which appears in the dispersion relation, unlike in the nonrelativistic theory where $\nu = \hbar/2m$.} and the drift is
\BE
b_\bx^\Psi(\phi;t) = \frac{c^2}{a^d} \Big( \frac{\del R(\phi;t)}{\del \phi_\bx} + \frac{\del S(\phi;t)}{\del \phi_\bx} \Big) =  \frac{\hbar c^2}{a^d}\big(\re + \im \big) \Big[\frac{1}{\Psi(\phi;t)} \frac{\del \Psi(\phi;t)}{\del \phi_\bx} \Big],
\EE
with $\Psi = \exp[(R+iS)/\hbar]$. Again, we are taking the perspective of Bohm and Hiley, where we assume the wave functional $\Psi(\phi;t)$ satisfies the Schr\"odinger equation, eq. (\ref{Schro}), and regard the process above as a stochastic guiding law, which generalizes that of Bohm \cite{Bohm-II:1951}, where only the $\del S/\del \phi$ term was included. Thus, the guiding law determines the evolution of the field beable from one time to the next, but this evolution is subject to random perturbations by a background noise field $\eta_{\bx}(t)$. The probability density $P(\phi;t)$ of field configurations approaches the quantum equilibrium limit $P(\phi,t) \to \rho(\phi;t) = |\Psi(\phi;t)|^2$, independently of the initial condition, at which point it satisfies the continuity equation,
\BE
\del_t \rho(\phi;t) + \frac{c^2}{a^d} \sumop_\bx \frac{\del}{\del \phi_\bx} \Big[\rho(\phi;t) \frac{\del S(\phi;t)}{\del \phi_\bx} \Big] = 0.
\EE
We emphasize that $P(\phi;t)$ denotes the probability (density) that the actual configuration of the field beable on a spatial lattice at time $t$ is $\phi = \{\phi_\bx\}$.

We discretize the process in \eqn{field-langevin} by sending $\eta_\bx(t) \to \sqrt{2\nu/a^d \eps} \; \what \eta_{t,\bx}$, where $\mrm{E}[\what \eta_{t,\bx}\what \eta_{s,\by}] = \delta_{t,s} \delta_{\bx,\by}$, to obtain a discrete Euler step
\BE
\phi_{t+\eps,\bx} = \phi_{t,\bx} + \eps b_\bx^\Psi(\phi_t,t) + \sqrt{2\nu\eps/a^d} \; \what{\eta}_{t,\bx}.
\EE
As in the case of hydrogen, here we limit ourselves to the simple Euler step, but note that Runge-Kutta methods exist also for lattice fields \cite{Batrouni}. Now, the natural dimensionless field variable in simulations is $\phi_\bx/\sqrt{\hbar c a^{1-d}}$, and time is measured in units of $\tau_0 = a/c$, which therefore requires knowledge of the lattice spacing $a$ to specify in physical units. $a$, in turn, is determined by ``setting the scale'' of the simulation \cite{Montvay}. For a free theory, since the bare and renormalized masses coincide, and the dimensionless input parameter $M := amc/\hbar$ is the bare mass, one need only set $m$ equal to a known particle mass, e.g., the pion with $m \approx 140 \; \mrm{MeV}/c^2$, which then implies $a = \hbar M / mc \approx 10^{-15}$ meters and $\tau_0 \approx 10^{-23}$ sec, when $M \approx 1$. For an interacting theory, one needs to measure the dimensionless renormalized mass $M_R$ from, e.g., the exponential decay of a 2-point correlation function, and then set $a = \hbar M_R / mc$, where $m$ is a chosen physical mass, like that of the pion. We go back to natural units $\hbar = c = 1$ in what follows.

\subsection{Free Scalar Fields}

\begin{figure}
\centering
\begin{subfigure}{.5\textwidth}
  \centering
  \includegraphics[width=1\linewidth]{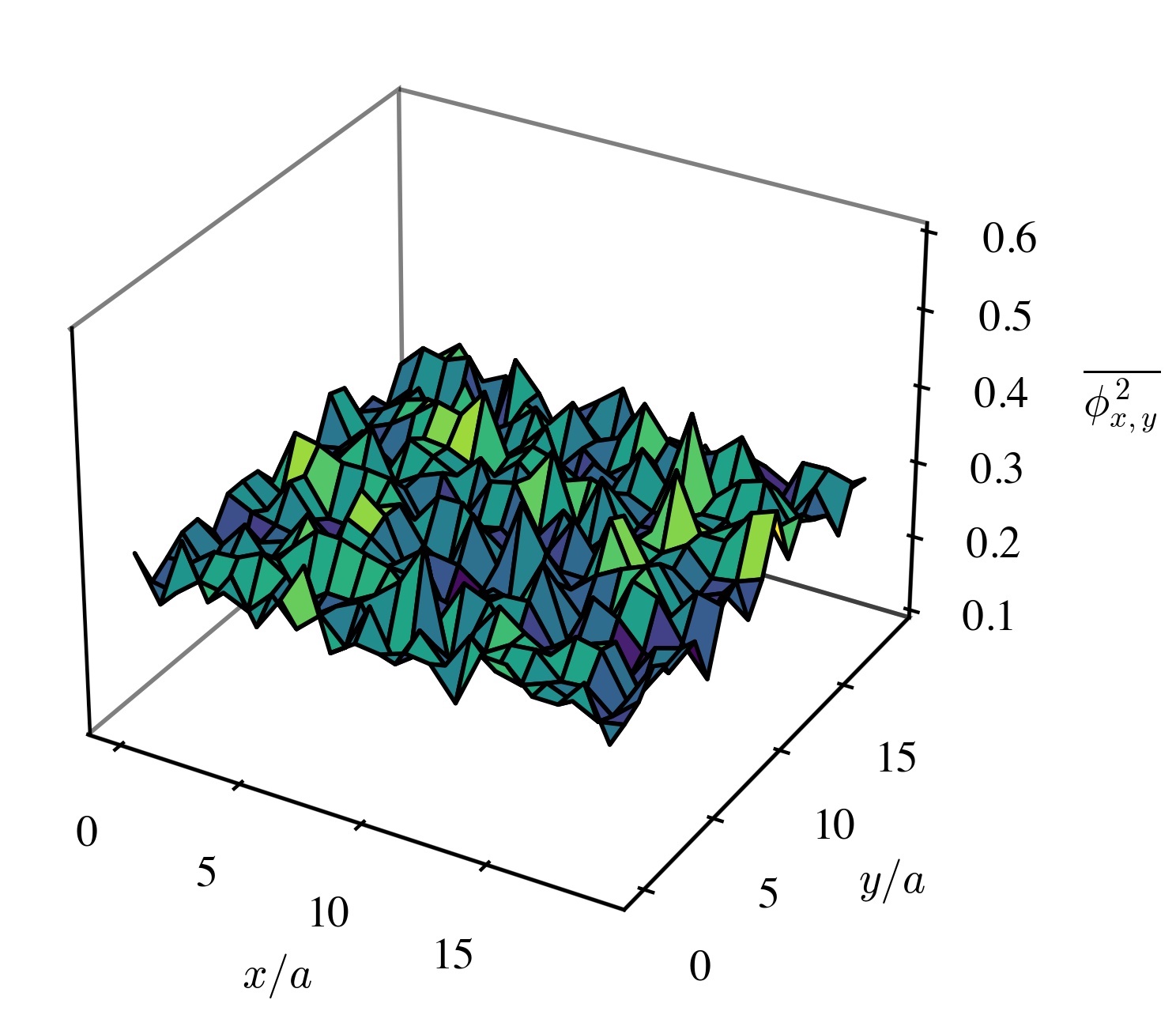}
  \caption{}
  %\label{fig:sub1}
\end{subfigure}%
\begin{subfigure}{.5\textwidth}
  \centering
  \includegraphics[width=1\linewidth]{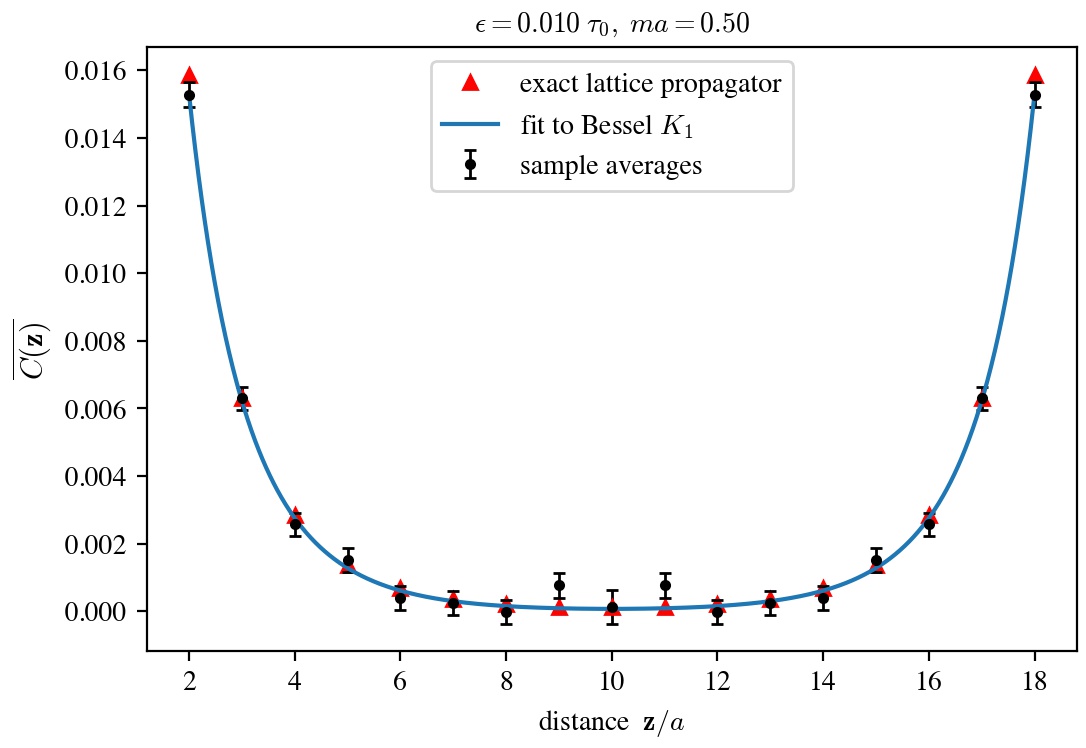}
  \caption{}
  %\label{fig:sub2}
\end{subfigure}
\caption{\small{(a) Average over 100 sample configurations of the square of the field beable determined by the  free field vacuum state $\Psi_0(\phi)$ on a $20\times 20$ periodic plane. (b) Correlation function $C(\bo z) = \overline{\phi_{\bx+\bo z} \phi_\bx}$ computed from 100 sample configurations such as in (a), compared with the exact result.}}
\label{fig:vacuum}
\end{figure}

As a first example of a stochastic field beable evolving in accordance with Nelson's guiding law, we begin with the free-field vacuum state $\Psi_0(\phi;t) = \me^{-iE_0t} \Psi_0(\phi)$, where $\Psi_0$ is given in eq. (\ref{free-vacuum}). The Langevin equation reads
\BE
\del_t \phi_\bx(t) = - a^d \sumop_{\by} \omega_{\bx,\by} \phi_\by(t) + \eta_{\bx}(t),
\EE
where $\omega_{\bx,\by} = (1/V)\sumop_{\bp} \me^{i\bp(\bx-\by)} \omega_\bp$.\footnote{The \emph{square} of $\omega_{\bx,\by}$ as a matrix 
is (minus) the lattice Laplacian operator, plus a mass term.} This process is formally akin to the Ornstein-Uhlenbeck process, and due to its simplicity, the solution can be written out explicitly:
\BE \label{OUsol}
\phi_\bx(t) = \sumop_{\by} K_{\bx,\by}(t) \; \varphi_\by + \int_0^t \!\dd s \sumop_\by K_{\bx,\by}(t-s) \; \eta_{\by}(s),
\EE
where $\varphi_\bx = \phi_\bx(0)$ and $K_{\bx,\by}(t)$ is a sort of diffusion kernel,
\BE \label{rel-kernel}
K_{\bx,\by}(t) = \frac{1}{V} \sumop_{\bp} \me^{i\bp (\bx-\by) - \omega_\bp t} ,
\EE
which differs from the standard lattice heat kernel by having $\omega_\bp = \sqrt{\hat{\bp}^2+m^2}$ in the exponent rather than $\hat{\bp}^2$. The exponential decay of the first term in eq. (\ref{OUsol}) characterizes the relaxation toward quantum equilibrium, which in this case, is time-independent --- though the field beable is always fluctuating wildly. In Figure \ref{fig:vacuum} (a) we display an average over 100 sample field configurations at some time $t \gg t_\mrm{eq}$ (which we determine ``by-eye''). Although visually the field is a jumble of noise, the statistics of the field matches that of the standard free-field vacuum state. In Figure \ref{fig:vacuum} (b), for example, we plot the spacelike correlation function $D_{\bx,\by} = \langle\phi_\bx \phi_\by\rangle_{\Psi_0}$ computed as an average over 100 such field configurations, comparing it with the exact form. The picture of the field beable determined by the vacuum state $\Psi_0$ as a fluctuating, dynamical quantity, is the field theory analogue of what was observed in the hydrogen atom. Whereas in the Bohmian case the field is static, or ``frozen'' across space, in the stochastic theory it is an entity which fluctuates about in dynamic equilibrium: the drift term wants to drive the momentum modes toward zero (toward spatially-constant configurations), but they are prevented from settling at zero by the random noise.

\begin{figure}
\centering
\begin{subfigure}{.5\textwidth}
  \centering
  \includegraphics[width=1\linewidth]{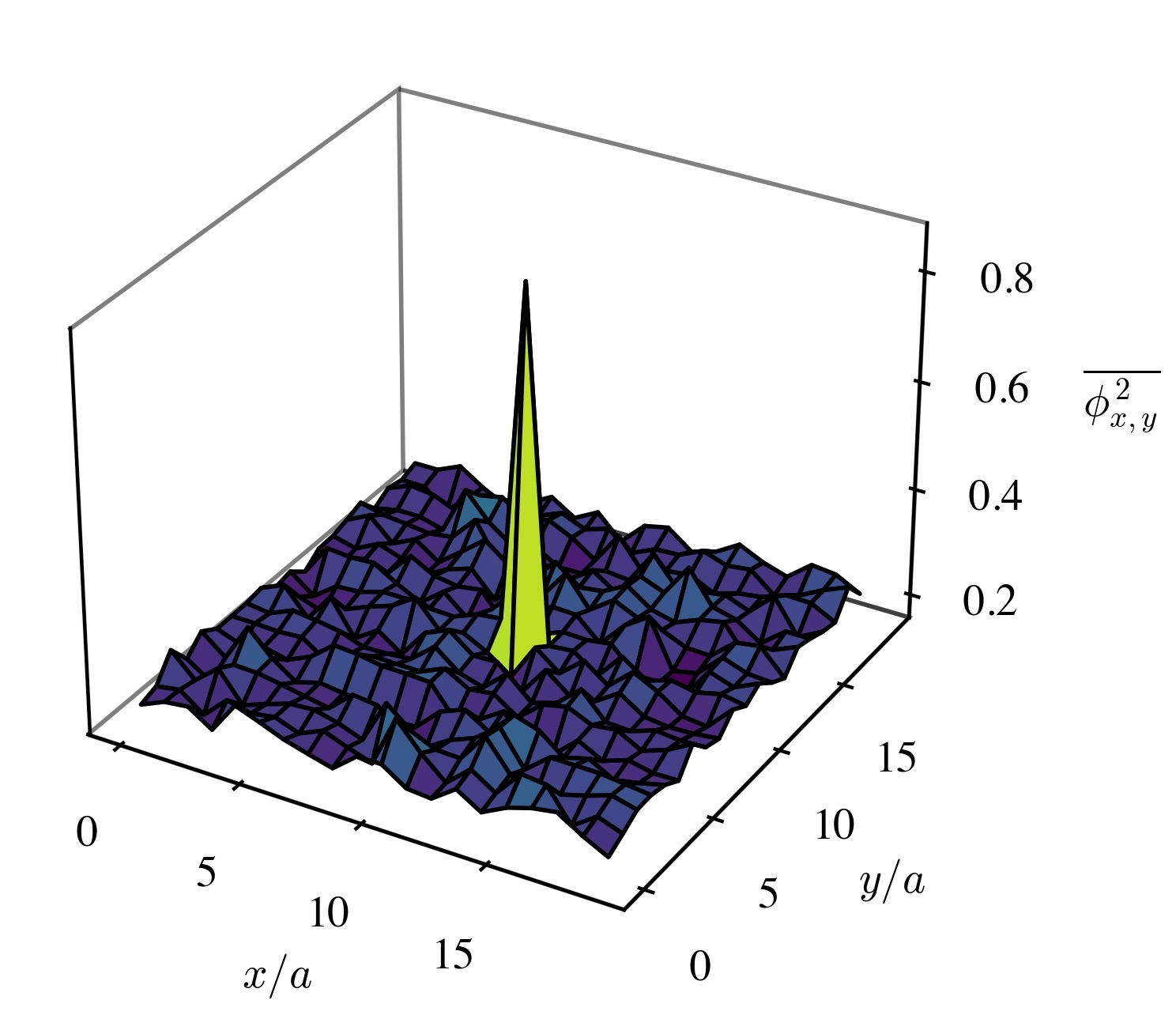}
  \caption{}
  %\label{fig:sub1}
\end{subfigure}%
\begin{subfigure}{.5\textwidth}
  \centering
  \includegraphics[width=\linewidth]{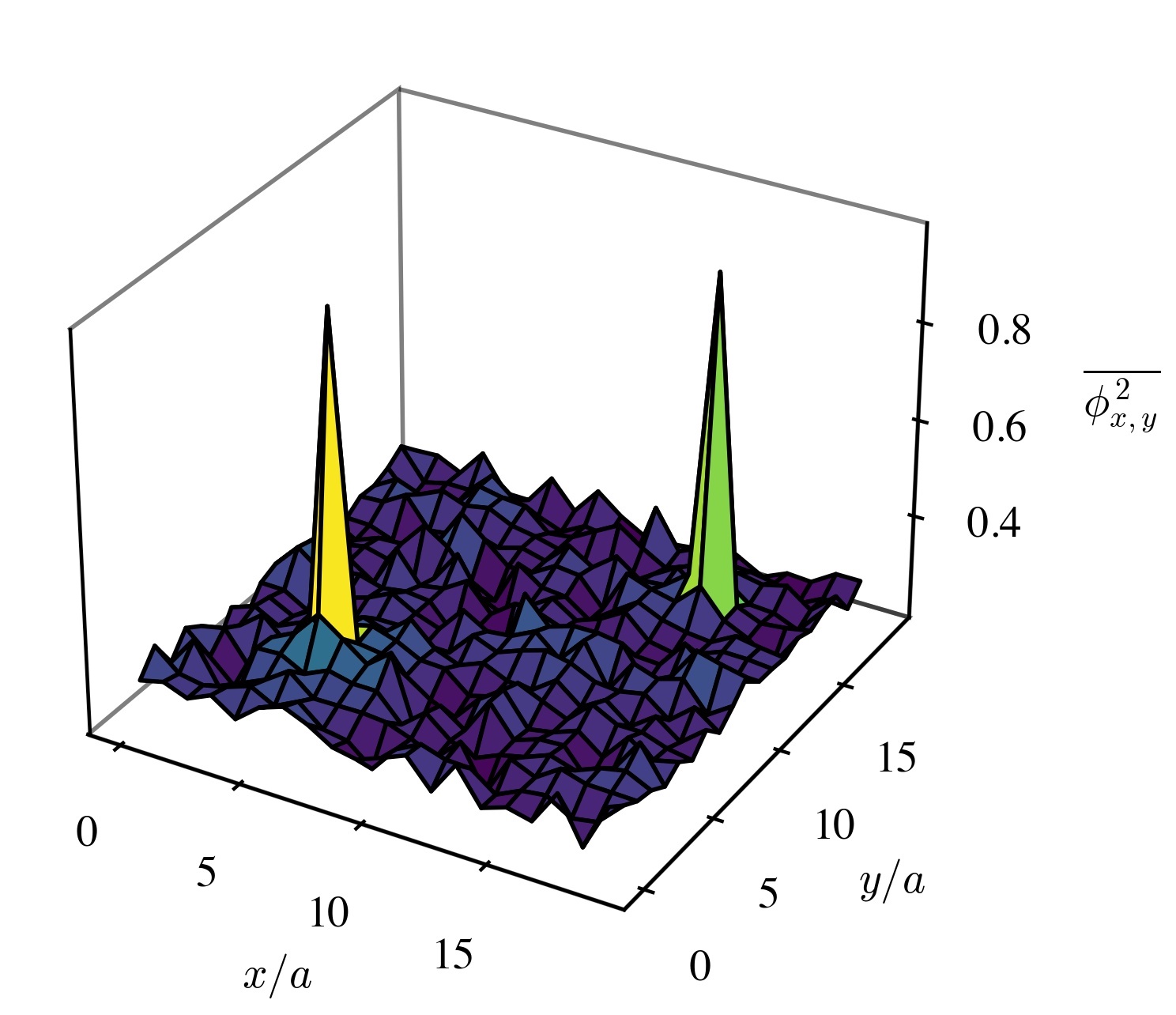}
  \caption{}
  %\label{fig:sub2}
\end{subfigure}
\caption{\small{Average over 100 samples of the field beable corresponding to the 1-particle (a) and 2-particle (b) free field wave functionals $\Psi_1(\phi;\bx_1), \; \Psi_2(\phi;\bx_1,\bx_2)$. The locations of the spikes coincide with the coordinates $\bx_1, \; \bx_2$. We took $am=0.8$ and simulated on a $20 \times 20$ lattice.}}
\label{fig:1p2p}
\end{figure}

Further observations of interest can be made by simulating the Nelsonian dynamics of $n$-particle static free-field states. By ``static,'' we mean that the states are independent of time. As such, we are for now merely using Nelson's guiding law to produce typical field configurations implied by certain wave functionals. A genuinely time-dependent state will be considered later in this section. Now, since the 1- and 2-particle states we found above were states of definite momentum, we consider instead the following states in position space,\footnote{These states are similar to the Newton-Wigner states considered, for example, in \cite{Myrvold:2015}, except for the particular momentum-dependence of coefficients. The states I use here are built from ``positive-frequency'' components of the field operator, $(\phi^+_\bx)^\dag$.}
\begin{align} \label{1p2p-states}
\Psi_1(\phi;\bx_1) & = \frac{1}{V} \sumop_\bp \frac{\me^{-i\bp \bx_1}}{2\omega_\bp} \; \what a_\bp^\dag \Psi_0(\phi) = \phi_{\bx_1}\Psi_0(\phi), \\
\Psi_2(\phi;\bx_1,\bx_2) & = \frac{1}{V^2} \sumop_{\bp_1,\bp_2} \frac{\me^{-i\bp_1 \bx_1 - i \bp_2 \bx_2}}{2\omega_{\bp_1}2\omega_{\bp_2}} \; \what a_{\bp_1}^\dag \what a_{\bp_2}^\dag \Psi_0(\phi) \nn\\
& = \big( \phi_{\bx_1} \phi_{\bx_2} - D_{\bx_1,\bx_2} \big) \Psi_0(\phi),
\end{align}
up to normalization constants.\footnote{Note that the normalization of $\Psi_1$ is $(D_{0,0})^{-1/2}$, implying the overlaps $\langle \Psi_1(\bx_1)|\Psi_1(\bx_2)\rangle = D_{0,0}^{-1} D_{\bx_1,\bx_2}$. Since $D_{0,0}$ is UV-sensitive, while $D_{\bx,\by}$ remains finite in the continuum limit, the overlap of normalized states is suppressed by the cutoff scale for non-coincident points.}
The drift vector for the 1-particle state is therefore
\BE
b_\bx(\phi;t) = \frac{\delta_{\bx,\bx_1}}{a^d \phi_{\bx_1}} -  a^d \sumop_{\by} \omega_{\bx,\by} \phi_\by.
\EE
The first term is due to the ``excitation'' by $\phi_{\bx_1}$; the second term is the contribution from the vacuum. We observe that a node exists at configurations $\phi$ with $\phi_{\bx_1} = 0$, analogous to the node we observed in the excited hydrogen eigenstate. Thus the configurations with $\phi_{\bx_1} = 0$ are ``repelled'' in the space of fields explored by the guiding law, which implies that typical field beables will be non-zero at $\bx_1$, where a spike, or lump (see below), tends to be present.\footnote{It's nevertheless true that $\langle \phi_{\bx} \rangle = 0$ in the state $\Psi_1$, since the ground state is symmetric under $\phi \to -\phi$. But whereas the configuration $\phi_\bx = 0$ is typical in the ground state, it is not typical in $\Psi_1$.} This spike is a \emph{regularity} in the field beable, and may not be manifested in a given particular configuration at some particular time; as such, it may be difficult to see in a single snapshot of a beable, especially given the presence of the random noise. Its structure is, however, brought out clearly in averages over an ensemble of beables, or in time averages along a single evolution. In Figure \ref{fig:1p2p} (a) we plot an average over 100 sample trajectories of $\phi^2$, at some time $t$ beyond the relaxation time.

It is clear that any state of the form $\phi_{\bx_1} \cdots \phi_{\bx_n} \Psi_0(\phi)$ will, along the same lines, yield a field with spikes at the locations $\{\bx_i\}$. But this remains true even of states like $\Psi_2$ above: the drift vector in the case of the 2-particle state, \eqn{1p2p-states}, comes out to
\BE
b_\bx(\phi;t) = \frac{1}{a^d} \frac{\delta_{\bx,\bx_1}\phi_{\bx_2} + \delta_{\bx,\bx_2} \phi_{\bx_1}}{\phi_{\bx_1}\phi_{\bx_2} - D_{\bx_1,\bx_2}} -  a^d \sumop_{\by} \omega_{\bx,\by} \phi_\by.
\EE
This drift repels configurations with $\phi_{\bx_1}\phi_{\bx_2} = D_{\bx_1,\bx_2}$. If the separation $\bx_1-\bx_2$ is large, $D_{\bx_1,\bx_2} \approx 0$, so simultaneously small field values at both sites is disfavored; this encourages lump structures at $\bx_1,\bx_2$. Figure \ref{fig:1p2p} (b) displays an example of such a field, where again we plot the average over 100 samples of $|\phi_\bx|^2$, observing clearly two spikes in the field.\footnote{See \cite{vid-2} for an animation of the field beable in this state.} We remark that, since the time step of the integrator is finite, it is possible to hit arbitrarily close to the nodes of the wave functional --- implying an enormous jump in the value of $\phi$ at $\bx_1$ or $\bx_2$ --- but again the frequency of the anomalous jumps is empirically observed to diminish as $\eps \to 0$, similar to the behavior in the $\psi_{200}$ case of Section 2. 

Away from the continuum limit, the ``particle'' structures in the field are not exactly point-like.
Indeed, for states of the form $\Psi_1 = a^d \sumop_\bx \psi_\bx \phi_\bx \Psi_0$, where $\psi_\bx$ is possibly time-dependent, the most probable field configuration $\varphi$ at time $t$ satisfies
\BE \label{probable-config}
a^d\sumop_\by \omega_{\bx,\by} \varphi_\by = \mrm{Re} \Big[ \frac{\psi_\bx}{a^d\sumop_\by \psi_\by \varphi_\by} \Big],
\EE
which was also observed by Valentini \cite{Valentini-Cushing} in the nonrelativistic limit $\omega_\bp \approx m$; he found that $\varphi$ ``mimics'' the function $\psi$, since  $\varphi_\bx(t) = \pm\psi_\bx(t)/\sqrt{m}$, for $\psi$ normalized and real. Away from the nonrelativistic limit, and for a point-source $\psi_\bx = a^{-d}\delta_{\bx,\bx_0}$, eq. (\ref{probable-config}) implies
\BE \label{config-ptsource}
\varphi_\bx = \frac{1}{V} \sumop_\bp \frac{\me^{i\bp(\bx-\bx_0)}}{\omega_\bp} \frac{1}{\varphi_{\bx_0}} = \frac{2D_{\bx,\bx_0}}{\varphi_{\bx_0}}.
\EE
Setting $\bx=\bx_0$ determines $\varphi_{\bx_0}^2 = (1/V)\sumop_\bp 1/\omega_\bp = 2 D_{0,0}$, and $D_{\bx,\bx_0}$ decays rapidly with distance from $\bx_0$ for $am > 0$. In fact, it is just the behavior displayed in Figure \ref{fig:vacuum} (b), which is fit well by Bessel $K_1$. In the continuum limit, however, $\varphi_\bx$ approaches a delta-like function, since $D_{\bx,\bx_0}$ has a finite continuum limit for fixed $\bx-\bx_0 \neq 0$, but $D_{0,0} \sim a^{-(d-1)}$ is UV-sensitive. Given that the continuum theory must also be regulated in some way,\footnote{For example, regularizing the canonical commutators by defining $[\what \phi(\bx), \what \pi(\by)] = if_\Lambda(\bx-\by)$ (with sufficiently quickly decaying $f_\Lambda$) rather than $i\delta(\bx-\by)$ yields a vacuum $\Psi_0(\phi) \propto \exp[-\frac{1}{2} \int \dd^d p \; \omega(\bp) \phi(\bp) \phi(-\bp)/ f_\Lambda(\bp)]$, which in turn yields a likeliest field configuration with $\varphi(\bx_0)^2 \propto \int \dd^d p f_\Lambda(\bp)/\omega(\bp)$, and is finite if $f_\Lambda(\bp)$ decays more quickly than $1/\bp^{d-1}$.} with a smooth (momentum) cutoff $\Lambda$, $\varphi_\bx$ is then expected to approach a \emph{smoothed} delta function as $a \to 0$. At distances much larger than the inverse cutoff $\Lambda^{-1}$, ultimately, $\varphi$ will \emph{look} like a localized spike.

\begin{figure}
\centering
\begin{subfigure}{.5\textwidth}
  \centering
  \includegraphics[width=1\linewidth]{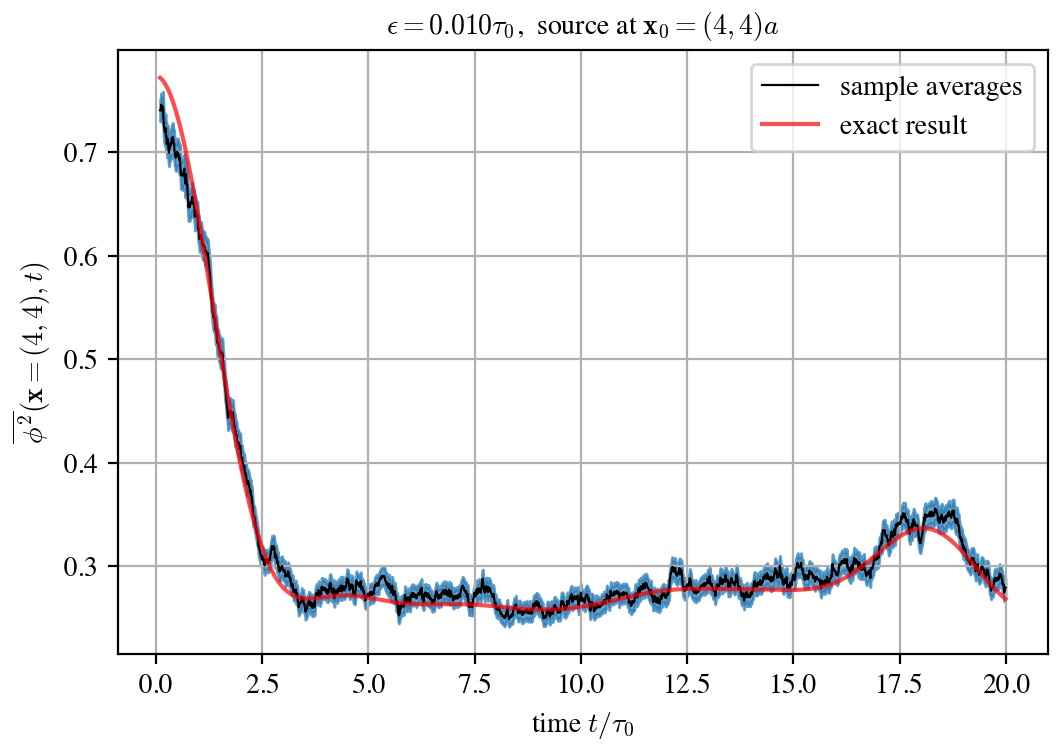}
  \caption{}
  %\label{fig:sub1}
\end{subfigure}%
\begin{subfigure}{.5\textwidth}
  \centering
  \includegraphics[width=\linewidth]{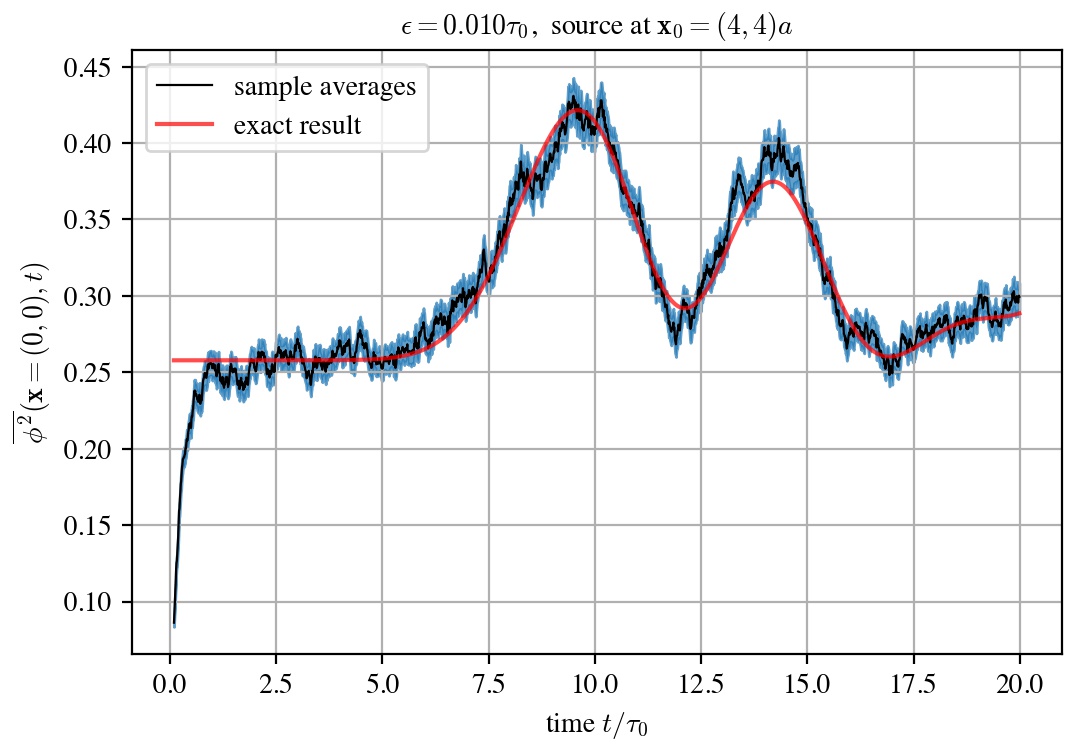}
  \caption{}
  %\label{fig:sub2}
\end{subfigure}
\caption{\small{Averages of $\overline{\phi^2_{\bx}}(t)$ over 2000 samples of the field beable corresponding to the time-dependent 1-particle state $\Psi(\phi,t;\bx_0)$ in \eqref{1ptdep}. Figure (a) looks at $\bx = (4,4)a = \bx_0$, where $\bx_0$ is the location of the source at $t=0$, while (b) looks at $\bx = (0,0)a$, the origin of the lattice. Blue bands denote the error on the mean. Red curves are the exact result from evaluating $\langle \Psi_t | \what \phi^{\;2}_\bx |\Psi_t\rangle$. The lattice size was $8\times 8$, and the mass was $am = 0.8$.}}
\label{fig:tdep-stats}
\end{figure}

We may also confirm the ability of Nelsonian QFT to reproduce correct statistics for genuinely time-dependent solutions of the Schr\"odinger equation, with nontrivial phases $S(\phi,t)$. As an example, we define the time-evolved 1-particle state (up to normalization)
\BE \label{1ptdep-state}
\Psi(\phi,t;\bx_1) := \frac{1}{V} \sumop_\bp \frac{\me^{-i\bp \bx_1}}{ 2\omega_\bp} \; \langle \phi | U_t \what a_\bp^\dag | 0 \rangle = a^d \sumop_\by K_{\bx_1,\by}(it) \; \phi_\by \Psi_0(\phi,t),
\EE
where $K_{\bx,\by}(it)$ is a relativistic free particle kernel, eq. \eqref{rel-kernel}, with $t\to it$. In this case, the drift vector is
\BE \label{1ptdep}
b^\Psi_\bx(\phi,t) = (\mrm{Re} + \mrm{Im}) \Big[\frac{K_{\bx_1,\bx}(it)}{a^d\sumop_\by K_{\bx_1,\by}(it) \; \phi_\by}\Big] - a^d \sumop_\by \omega_{\bx,\by} \phi_\by.
\EE
In Figure \ref{fig:tdep-stats} we plot the time-dependence of sample averages of $\phi^2_{\bx}$ at two locations, one at the source position $\bx = \bx_0=(4,4)a$ and another at the origin $\bx = (0,0)a$, together with the corresponding exact result,
\BE \label{1ptdep-evalue}
\langle \Psi_t | \what \phi^{\;2}_\bx |\Psi_t\rangle = D_{0,0} + 2 D_{0,0}^{-1}|D_{\bx,\bx_1}(t)|^2,
\EE
where $D_{\bx,\by}(t):=a^d\sumop_{\bxp} K_{\bx,\bxp}(it) D_{\bxp,\by}$ is the amplitude for propagation from $\by$ at $t=0$ to $\bx$ at $t$, finding good agreement.
The coefficient ``wave function'' in this state, $\psi_{\bx}(t) = K_{\bx,\bx_1}(it)$, is commonly taken to describe the probability density for an ensemble of particles propagating freely away from the initial source point $\bx_1$ in any direction (up to the breaking of rotation invariance by the lattice). The smooth diffusive behavior of $\psi_{\bx}(t)$ therefore appears to be displayed in the field beable (at the level of ensemble averages), as one might expect from the reflections on likeliest configurations above. Whether the diffusive behavior is a result of averaging over particulate-spikes such as in Figure \ref{fig:1p2p}, propagating away from the source point in different directions for different field beables, is not easy to verify, due to the noisiness of individual field configurations. The corresponding beable in the Bohmian case (where osmotic and noise terms are neglected) also exhibits diffusive behavior.\footnote{See \cite{vid-3} for an animation of this field.} Thus, it appears that beables corresponding to time-dependent $n$-particle states need not have spike structures in them, in general.

Before moving to the next section, let us remark upon the problem of ontology for wave \emph{functions}. In nonrelativistic QM, an $n$-body wave function propagates on a high-dimensional configuration space $\bb R^{nd}$, as opposed to the ordinary $\bb R^d$ of everyday experience, and this naturally raises concerns about whether it should be taken seriously as a real entity. In QFT, on the other hand, $n$-body wave functions are a derived concept, although there is not complete agreement on how to extract them from arbitrary quantum states $\ket \Psi$. Their definition has roughly the form
\BE
\psi^{(n)}_{\bx_1,...,\bx_n}(t) := \langle n;\bx_1,...,\bx_n| \Psi(t) \rangle,
\EE
where $\ket{n;\bx_1,...,\bx_n}$ is a product of ``localized'' operators, such as $\what \phi_{\bx}$, acting on the state $\ket 0$. For example, taking $\what \phi_\bx$ as the local operator, the one-particle state, eq. (\ref{1ptdep-state}), has a corresponding wave function
\BE
\psi_\bx(t) = a^d \sumop_\by K_{\bx_1,\by}(it) D_{\by,\bx} = D_{\bx_1,\bx}(t).
\EE
In a sense, therefore, because the wave functions are derivative from the wave functionals, there is no problem of ontology for the $n$-body wave functions, though of course there remains the problem of wave \emph{functional} ontology. Here, we wish only to elaborate on the derivative aspect of the $\psi^{(n)}$. Namely, if the beable of the theory is a field $\phi_\bx(t)$ across ordinary space $\bb R^d$, how does the ``many-body-ness'' of $\psi^{(n)}$ emerge?

First, note that the \emph{likeliest} field configuration, $\varphi$, for an arbitrary $n$-body wave functional
$\Psi_n(\phi,t) = \mcal P_n(\phi,t) \Psi_0(\phi)$,
with $\mcal P_n$ polynomial in $\phi$, is determined by a generalization of eq. (\ref{probable-config}):
\BE
\varphi_\bx = a^d \sumop_{\by} 2 D_{\bx,\by} \; \mrm{Re} \Big[ \frac{1}{\mcal P_n(\phi,t)} \frac{\del \mcal P_n(\phi,t)}{\del \phi_\by} \Big] \Big|_{\phi = \varphi} .
\EE
If $\mcal P_n(\phi,t) = \sumop_\by \psi_{\by_1,...,\by_n} \phi_{\by_1} \cdots \phi_{\by_n}$, and $\psi$ depends on $n$ ``source coordinates'' $\bx_i$, then the likeliest configuration $\varphi_\bx$ also depends on the source coordinates $\bx_i$, similar to what we observed in eq. (\ref{config-ptsource}), and made clear in Figure 5.\footnote{This observation is not limited to just Bohm or Nelson's theories: it applies to any interpretation which takes wave functionals and field beables seriously.} This is true also for the beables determined by the guiding law: because $\Psi_n$ depends on the source coordinates, solutions of the guiding law can be written as
\BE
\phi_\bx(t) = \mPhi_t(\bx,\bx_1,\dots,\bx_n) ,
\EE
where $\mPhi_t$ is a function on $\bb R^{(n+1)d}$. Notice that the value of $\phi$ at a point $\bx$ in physical (everyday) space therefore depends on $n$ ``background'' coordinates $\bx_i$ having to do with the initial source coordinates in the quantum state, and which, roughly speaking, correspond to $n$ initial particle coordinates. Now, since the $n$-body wave functions were embedded into the state $\Psi$, they are also embedded into the beable $\phi_\bx(t)$, and the high-dimensionality of the $\psi^{(n)}$ is manifested only by the presence of dependence on the background coordinates in $\mPhi_t$. In the time-independent 2-particle state, for example, there are two spikes in the field located at the two source coordinates. In the time-dependent version, the two coordinates are initial points from which waves in the field propagate outward (see footnote 25). The functions $\psi^{(n)}$ can be extracted from $n$-point expectation values over an ensemble of field beables, as in eq. (\ref{1ptdep-evalue}). This behavior is present for both Nelsonian and Bohmian theories of the bosonic field.\footnote{See \cite{vid-4} for an example of a Bohmian field beable for a 2-particle time-dependent state.}

\subsection{Interacting Scalars} 

In the case of scalar fields with self-interaction, e.g., $\phi^4$, the exact solution to the Schr\"odinger equation is not known, but various perturbative methods exist for studying such systems. We may therefore get some insight into the behavior of the Nelsonian field beable for an interacting theory by studying the guiding law under certain approximations of the wave functional. To that end, we adopt the perturbative strategy outlined in Hatfield \cite{Hatfield} for the Schr\"odinger picture, which we now review and slightly extend.

Begin by writing the ground state wave functional as $\Psi_0(\phi,t) = \exp[R(\phi)+iS(\phi,t)]$, assuming that the phase $S(\phi,t)=-E_0t$ is independent of $\phi$. The Schr\"odinger equation (in momentum space) for $R(\phi)$ comes out to
\begin{align}
E_0 = -\frac{V}{2}\sumop_\bp & \Big[ \frac{\del^2 R}{\del \phi_\bp \del \phi_{-\bp}} + \frac{\del R}{\del \phi_{\bp}} \frac{\del R}{\del \phi_{-\bp}} \Big] \nn\\
& + \frac{1}{2V}  \sumop_\bp \omega^2_\bp \phi_\bp \phi_{-\bp} + \frac{\lambda}{4! V^3} \sumop_{\{\bp_i\}} \delta(\bp_\mrm{tot}) \; \phi_{\bp_1} \phi_{\bp_2} \phi_{\bp_3} \phi_{\bp_4}.
\end{align}
We assume a perturbative ansatz, $R=R^{(0)}+R^{(1)}+\cdots, \; E_0 = E_0^{(0)} + E_0^{(1)}+\cdots$, with $R^{(k)}, E_0^{(k)}$ of order $k$ in $\lambda$. $R^{(0)}$ is just the exponent of the free-field vacuum in eq. (\ref{free-vacuum}), which is  quadratic in $\phi$. One finds that a polynomial of order four in $\phi$ solves the order-$\lambda^1$ equation, with the result
\begin{align}
R^{(1)}(\phi) = -\frac{\lambda}{4 V} & \sumop_{\bp} \Big(\frac{1}{2\omega_\bp V} \sumop_\bk \frac{1}{2\omega_\bp + 2 \omega_\bk} \Big) \phi_{\bp} \phi_{-\bp} \nn\\ 
& - \frac{\lambda}{4! V^3} \sumop_{\{\bp_i\}} \frac{\delta(\bp_\mrm{tot})}{\sumop_{i=1}^4 \omega_{\bp_i}} \; \phi_{\bp_1} \phi_{\bp_2}\phi_{\bp_3} \phi_{\bp_4}, \\
E_0^{(1)} = \frac{\lambda}{8 V} \Big(& \sumop_{\bp}\frac{1}{2\omega_\bp} \Big)^2.
\end{align}
Hence the perturbative $\phi^4$ vacuum wave functional is of the form
\BE
\Psi_0(\phi) = \exp \Big( -\frac{1}{2V} \sumop_\bp \Omega_\bp \phi_{\bp} \phi_{-\bp} - \sumop_{\{\bp_i\}} g_{\{\bp_i\}} \phi_{\bp_1} \phi_{\bp_2}\phi_{\bp_3} \phi_{\bp_4} + O(\lambda^2\phi^6) \Big)
\EE
up to normalization, where the coefficients are
\begin{align} \label{vac-coeffs}
\Omega_\bp = \omega_\bp  + & \frac{\lambda}{4\omega_\bp V} \sumop_\bk \frac{1}{\omega_\bp + \omega_\bk} + O(\lambda^2), \\ %\approx \sqrt{\omega_\bp^2 + \Sigma_\bp} \\
g_{\bp_1 \bp_2 \bp_3 \bp_4} & = \frac{\lambda}{4! V^3} \frac{\delta(\bp_\mrm{tot})}{\sumop_{i=1}^4 \omega_{\bp_i}} + O(\lambda^2).
\end{align}
These coefficients are totally symmetric in their momentum indices. Similarly, one can go on to determine that the second order term $R^{(2)}$ is a polynomial of order six in $\phi$, and so on.\footnote{The coefficients have a structure analogous to the standard Feynman diagrams of QFT. The quartic term in $R^{(1)}$ corresponds to the tree-level diagram for the 4-point function, while the quadratic term in $R^{(1)}$ is the 1-loop correction to the 2-point function. The sextic term in $R^{(2)}$ gives a tree-level 6-point function; the quartic term gives the 1-loop contribution to the 4-point function (both 1PI and non-1PI diagrams); the quadratic term gives the 2-loop contribution to the 2-point function. The superficial degree of divergence of the loops is the same as the corresponding standard Feynman diagrams (see \cite{Amit} for a standard account). In this way we see how diagrams similar to those of standard QFT ``contribute'' to the vacuum wave functional.}

\begin{figure}
\centering
\begin{subfigure}{.5\textwidth}
  \centering
  \includegraphics[width=1\linewidth]{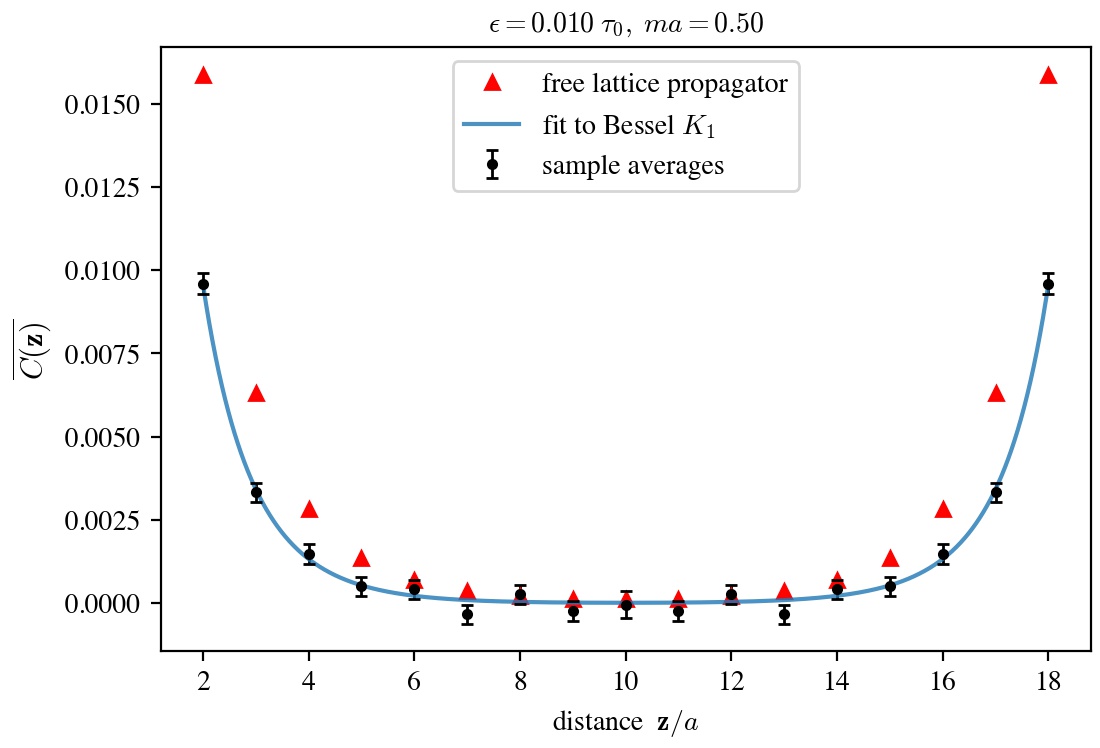}
  \caption{}
  %\label{fig:sub1}
\end{subfigure}%
\begin{subfigure}{.5\textwidth}
  \centering
  \includegraphics[width=0.9\linewidth]{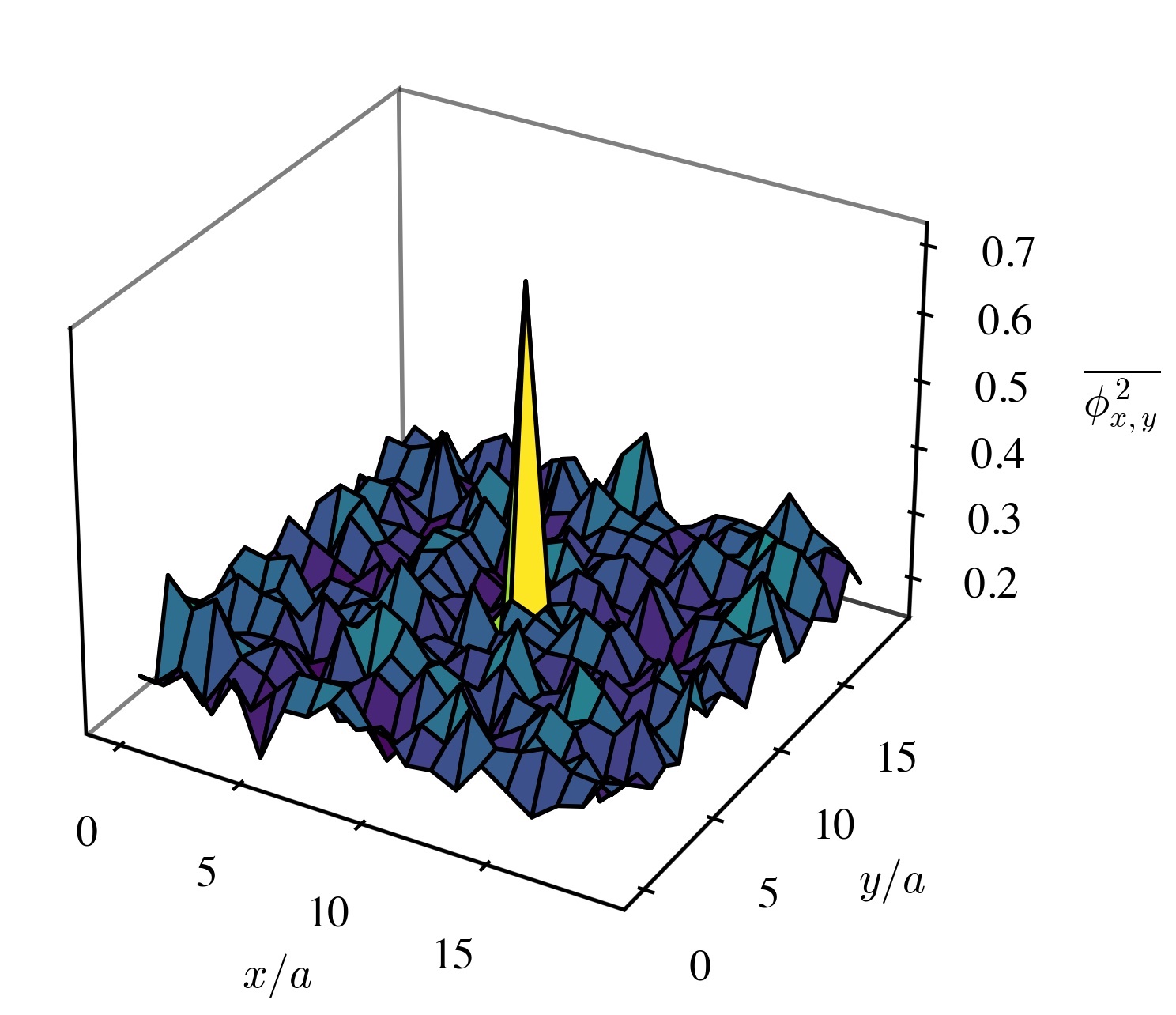}
  \caption{}
  %\label{fig:sub2}
\end{subfigure}
\caption{\small{(a) Average over 1000 sample configurations of correlations in the field beable generated by an approximation to the vacuum drift, \eqn{phi4-vac-drift-approx}, with bare coupling $a\lambda=1.0$ and bare mass $am=0.5$. The renormalized mass, obtained by fitting $C(\bo z)$ to a Bessel function $K_1(m_R z)$, was $am_R = 0.77(4)$ with $\chi^2/\mrm{dof}=0.51$. (b) Average over 100 configurations generated by the 1-particle drift, \eqn{1p-drift-int}, with $a\lambda = 4.0$, $am=0.8$, and $\eps=0.001\tau_0$.}}
\label{fig:vphi4-plots}
\end{figure}

The stochastic guiding law for the field beable, in a Nelsonian treatment of the vacuum state, then has drift
\BE \label{phi4-vac-drift}
b_\bx^{\Psi_0}(\phi;t) = - a^d \sumop_{\by} \Omega_{\bx,\by} \phi_\by - 4 a^{3d} \sumop_{\{\by_i\}} g_{\bx, \by_2, \by_3, \by_4} \phi_{\by_2}\phi_{\by_3} \phi_{\by_4} + O(\lambda^2),
\EE
where $\Omega_{\bx,\by}$ and $g_{\{\bx_i\}}$ are Fourier transforms of the corresponding momentum space functions, and are totally symmetric in their position indices.
To first order, then, the interaction has led to a cubic contribution to the drift, and a modified dispersion matrix $\Omega$. In principle, since we know $g$ explicitly, we could simulate the truncated process. However, the contraction of a rank-4 tensor like $g$ becomes expensive in a simulation, scaling with the volume like $V^4$, and so we resort to an approximation. To that end, we perform an expansion of $
\Omega_\bp$ and $g_{\{\bp_i\}}$ in momenta (i.e. a derivative expansion, in position space) and keep only the leading terms, resulting in
\begin{align} \label{phi4-vac-drift-approx}
b_\bx^{\Psi_0}(\phi) = - \sumop_{\by} & \omega_{\bx,\by} \phi_\by  - \frac{\lambda}{4m} I_1 \phi_\bx - \frac{\lambda}{4! m} \phi^3_\bx + O(\lambda \hat \Delta, \lambda^2),
\end{align}
where $I_1 = (1/V)\sumop_\bp 1/(m+\omega_\bp)$ and $\hat \Delta f_\bx = \sumop_\mu \hat \del^*_\mu \hat \del_\mu f_\bx$ is the lattice Laplacian. The second term on the first line, which contains a UV-sensitive integral, can be regarded as a shift in the mass parameter. In Figure \ref{fig:vphi4-plots} (a) we plot the correlation function from an ensemble of fields generated by \eqn{phi4-vac-drift-approx}. We find qualitatively similar behavior to the free-field vacuum, though the correlator is different now, as expected: the interaction leads to a shift in the physical mass with respect to the free theory. Also note that by simulating with a truncated drift function, we are not probing the true $\phi^4$ ground state; we will comment on the drift for the untruncated ground state later in this section.

The qualitative structure of $n$-particle states is also similar to the free case. To see this, we make the ansatz \cite{Haag-Coester}
\BE
\Psi_n(\phi; \bp_1,\dots,\bp_n) = \mcal P_n(\phi;\bp_1,\dots,\bp_n) \Psi_0(\phi),
\EE
where $\mcal P_n$ is a polynomial in $\phi$ with $n$ external momenta $\bp_i$. The Schr\"odinger equation implies that $\mcal P_n$ satisfies\footnote{Equations of striking similarity to these appear in the context of functional RG, where $\ln \Psi_0$ is replaced by an RG fixed point action, and $\mcal P_n$ is replaced by a scaling operator associated with that fixed point; see, e.g., \cite{Rosten:2010}. A connection between RG and stochastic processes has been explored in \cite{Carosso:2019,Carosso-Stoch}.}
\BE
(E_n - E_0) \mcal P_n(\phi) = - \frac{V}{2} \sumop_\bk \Big( \frac{\del^2 \mcal P_n(\phi)}{\del \phi_{\bk} \del \phi_{-\bk}} + 2 \frac{\del \mcal P_n(\phi)}{\del \phi_\bk} \frac{\del \ln \Psi_0(\phi)}{\del \phi_{-\bk}} \Big).
\EE
As an example, we compute $\mcal P_1(\phi)$ to first order in perturbation theory. One finds\footnote{There are also terms proportional to $\phi_{\pm\bp}$ with undetermined coefficients, reflecting the degeneracy of the free states $\ket{1,\pm\bp}$. We have set these coefficients to zero.}
\BE \label{phi4-1p-mom}
\mcal P_1(\phi;\bp) = C_\bp \phi_{-\bp} - \frac{\lambda C_\bp}{6 V^2} \sumop_{\{\bk_i\}} \frac{\delta(\bk_\mrm{tot} + \bp)}{\big(\sumop_i \omega_{\bk_i}\big)^2 - \omega_\bp^2} \;  \phi_{\bk_1}\phi_{\bk_2}\phi_{\bk_3} + O(\lambda^2),
\EE
where $C_\bp$ is the normalization of the free-field state. We divide by $C_\bp$ and go to position space via
\BE
\mcal P_1(\phi;\bx_1) := \sumop_{\bp} \frac{\me^{-i\bp \bx_1}}{C_\bp} \mcal P_1(\phi;\bp).
\EE
We approximate this state, as we did the ground state, by keeping only the leading term in its derivative expansion. The result is
\BE \label{1p-int-P}
\mcal P_1(\phi;\bx_1) = \phi_{\bx_1} - \frac{\lambda}{48 m^2} \phi^3_{\bx_1} + O(\lambda \hat \Delta, \lambda^2)
\EE
Interestingly, we see that not only configurations with $\phi_{\bx_1}=0$ are nodes, but also configurations with $\phi_{\bx_1}^2 = 48m^2/\lambda$ (in this approximation). Noting that the drift for any state of the form $\Psi_n = \mcal P_n \Psi_0$ can be written as (for $\mcal P_n$ real)
\BE \label{1p-drift-int}
b_\bx^{\Psi_n} (\phi) = \frac{1}{\mcal P_n(\phi)} \frac{\del \mcal P_n(\phi)}{\del \phi_\bx} + b_\bx^{\Psi_0}(\phi),
\EE
it is then simple to compute the drift for $\Psi_1(\phi;\bx_1)$.
Since $\phi_{\bx_1}=0$ is still a node of $\Psi_1$, as in the free case, the field tends to be non-zero at $\bx_1$, which yields a lump structure. In Figure \ref{fig:vphi4-plots} (b) we plot an average over 100 configurations of the beable generated by \eqn{1p-drift-int} with the truncation in \eqn{1p-int-P}, observing a spike in the field at the location $\bx_1$, similar to the free-field case. We expect this behavior to persist even without performing the derivative expansion, so long as the Fourier transform of the cubic term in \eqn{phi4-1p-mom} is sufficiently local, for then configurations that have $\phi_{\bx} \approx 0$ for $\bx$ near $\bx_1$ are still repelled in the guiding law.

We have been considering low-lying eigenstates of the interacting $\phi^4$ Hamiltonian to first order in $\lambda$. And we have seen that UV-sensitive integrals enter into the guiding law in this approximation, as in the constant $I_1$ in \eqn{phi4-vac-drift}; these integrals are similar to those appearing in standard field theory, and which, historically, led to the theory of renormalization \cite{Cao-Schweber,Amit}. Although the proofs of renormalizability were initially carried out in the interaction picture and the path integral representation, the renormalizability of Schr\"odinger picture QFT was established finally in 1981 by Symanzik \cite{Symanzik:1981}. Symanzik carried out his proof using a path integral representation of the ground state, however, and this slightly obscures the relationship to the type of Schr\"odinger picture calculations we have carried out here; a detailed analysis will be done in future works. The expected upshot is that through the definition of a renormalized field variable, $\phi_R = Z_1 \phi$, and a few renormalized parameters, including $m_R$ and $\lambda_R$, all expectation values of $\phi_R$ (without coincident points) will be insensitive to the UV cutoff, once expressed in terms of these parameters. A guiding law for $\phi_R$ may then be defined using the wave functional expressed in terms of the renormalized parameters, and the beables generated by the guiding law will exhibit the statistics of a distribution whose moments are UV-insensitive, since those moments are renormalized. The UV-insensitivity of observables involving $\phi_R$ implies that they will have finite continuum limits ($a \ll m_R^{-1}$).

It is also expected that an interacting theory will allow for particle creation and annihilation events. This is usually inferred by observing that the interaction potential $\mcal V$ couples Fock states of different particle number, or in the fact that $n\to m$ scattering amplitudes need not vanish for $n \neq m$. The detailed processes that occur at intermediate times, i.e. in the interaction region where the ``particles'' are not asymptotically separated, is left unaddressed by standard QFT, apart from a heuristic interpretation based on Feynman diagrams. In the Nelsonian (and Bohmian) frameworks, we find the possibility of giving detailed accounts of what might be going on during such processes, since the field beable will evolve according to whatever the wave functional happens to be. If a generic time-dependent wave functional involves a superposition of Fock states with differing particle number, then \emph{several} $n$-body wave functions (for different $n$) will be embedded in the field beable, along the lines discussed in the previous section, and the ``influence'' of different sectors at any particular time will be determined by relative (time-varying) weights in the superposition, similar to how the different terms in the hydrogen superposition, eq. (\ref{superposition}), influence the motion of the electron. Thus, it is conceivable that a beable initially with $n$ lumps may, upon introducing an interaction, evolve continuously into a field with $m$ lumps; but it remains to exhibit such states in more detail. Properly relating this behavior to what is observed in particle accelerators will furthermore require input from the Bohmian theory of measurement \cite{Bohm-II:1951,Cushing-Copenhagen,Nikolic:2009}.

Lastly, we have so far been dealing exclusively with bosonic field theory, and one naturally wonders how to incorporate fermionic fields in the Nelsonian framework.\footnote{The extension to (pure) lattice gauge theories should be possible, without too much difficulty, by utilizing methods from the stochastic quantization of such theories \cite{Batrouni,Damgaard}.} The difficulty lies in the traditional use of Grassmann numbers $\eta_\bx$ to represent fermion fields: it seems important that the wave functionals should be c-numbers, since one wants to interpret $|\Psi(\phi)|^2$ as giving a probability density (in quantum equilibrium, at least), whereas it is difficult to interpret a Grassmann-valued functional $|\Psi(\eta)|^2$ as a probability density \cite{Struyve:2007}. Without yet addressing this issue head-on, we remark that it should be possible to encode all fermionic aspects of a given QFT into the behavior of the bosonic fields that interact with them. For example, in lattice QCD, one never directly simulates with Grassmann variables: the fermionic part of the QCD action is integrated out of the path integral, resulting in an effective action for the gluons, and expectation values involving fermion operators are likewise evaluated in terms of the gluon field only. This strategy might be employable also in Nelsonian QFT, if one can find an effective guiding law for the bosonic fields, perhaps along the lines suggested in \cite{Struyve-Westman} for de Broglie-Bohm theory. We shall pursue this in future work.

\section{Discussion}

In this paper we have simulated Nelsonian QFT and found that it provides an intuitive way of assigning concrete processes, consisting in the stochastic evolution of a field beable, to the states (wave functionals) appearing in QFT. We may compare this strategy with other proposals in the literature for ontological interpretations of QFT. We focus on the Bohmian proposals, since that of Nelson is most similar to them in spirit. In particular, the Nelsonian account most closely resembles the standard de Broglie-Bohm (dBB) theory for bosonic QFT \cite{Bohm-II:1951,Bohm-Hiley,Holland,Struyve-Westman,Struyve:2007},\footnote{See also \cite{Derakhshani:2021} for a Lorentz covariant proposal.} except that it is stochastic and includes an osmotic term in the guiding law. On the relation between field-based dBB theory and that of Nelson, we make two observations.

First, it was remarked earlier that the Bohmian field corresponding to the Fock vacuum is static, or frozen in time. This remains true when including the full time-dependence of the state, since it only affects a multiplication by $\me^{-iE_0t}$ --- and it seems to remain true even for the vacuum state of the interacting $\phi^4$ theory. Unlike the hydrogen case, moreover, there does not seem to be an obvious analogue of the ``spin'' generalization that would render the behavior more plausible.
Echoing the sentiment of Bohm and Hiley quoted in the introduction, the stochastic picture of the vacuum state of a QFT seems to more closely reflect the intuitive picture of the ``quantum vacuum,'' namely, as a wildly fluctuating field. To be clear, having an intuitively satisfying picture does not force one to prefer Nelson's theory over the Bohmian theory; both theories have been argued to be empirically consistent with standard QFT. We also stress that both theories take the wave function(al) amplitude as input, on some level: in the initial conditions, for Bohmian mechanics, and in the osmotic term, for Nelsonian mechanics. 

Second, the particle-like behavior of certain field beables in the context of dBB QFT has been remarked upon in the past \cite{Valentini-Cushing,Kaloyerou-Cushing,Dewdney-Cushing}, though perhaps not fully explored. We have seen that the Nelsonian theory, however, makes it quite clear how certain time-independent $n$-particle wave functionals can generate, via the stochastic guiding law, field beables with particle-like characteristics. And although an individual beable for such states at a certain time looks quite noisy, the spike structures would be clear from a time average along that single beable's history. In the Bohmian case, by contrast, an individual beable need not display those characteristics, nor would it be dynamically produced in such a quantum state. For both Nelsonian and Bohmian accounts, however, the beables appear to have a diffusive character once one considers typical \emph{time-evolved} $n$-particle Fock states, rather than persistent spikes that move around in space, as might have been expected (or desired). The conditions under which the beable will retain \emph{stable} spike/lump structures throughout a realistic time evolution requires further study. 

Next, it should be noted that many of the recent proposals in Bohm-inspired QFT posit a fundamental particle ontology rather than a field ontology. In these models, the state of the universe at a given time (in some preferred Lorentz frame) consists in the specification of all particle positions, along with the wave function/al at that time. Some of these proposals are stochastic, while some are deterministic. In \cite{Bell-Beables,Durr:2004}, there is a probability of jumping, during any increment of time, to a configuration with differing particle number, with the probability being determined by the wave functional. In this case, particle creation processes involve a discrete change in the number of point-like particles: one particle might suddenly `pop' into two \cite{Vink:2017}. In other models, there is no fundamentally stochastic element to the theory, and particle creation is only in some sense ``apparent,'' as in the formation of an electron-hole pair out of the Dirac sea \cite{Colin:2007,Deckert:2019}, or Bohm-style measurements of particle number \cite{Nikolic:2009}. All of these models are elegant, and serve as concrete proposals for an ontological completion of QFT.

On the other hand, it is widely believed, according to the contemporary EFT philosophy, that most of the ``particles'' that enter our current QFTs are not fundamentally point-like objects, and instead have a finite effective size, being composed of perhaps smaller entities, themselves being of finite size. Although we cannot preclude the possibility of giving a reasonable interpretation of truly point-like particles in an EFT, there is an intuition among physicists that a point-like quality of particles goes hand-in-hand with an infinite UV cutoff (see, e.g., Bohm's comments on pg. 76 of  \cite{Quanta-Reality}).\footnote{This is perhaps borrowed from the association of the divergences in classical electromagnetism with a point-like model of electrons. In QFT, the association is more difficult to make, since the ontology of QFT is not as clear to begin with.} For these reasons, a field ontology seems to be more natural in the context of EFT, where finite-sized particles can be characterized as lumps in a field beable. We have seen that such a scenario is suggested by the Nelsonian account of QFT, albeit only for certain states. The field perspective also has the potential for incorporating particle creation and annihilation events in a natural way, as merely a continuous change of the lump structures in a single field beable.

Finally, there is a fundamental tension in the urge to provide an ontological completion of QFT, in light of the success of the EFT paradigm \cite{Williams}. According to this paradigm, one would expect that no given QFT will be completely reflective of the world: it will be accurate to some short distance scale, but it may break down beyond that scale. On the other hand, specifying an ontology seems to ``fill out'' the universe with certain objects and \emph{only} those objects, leaving no room for more detailed input from future theories. With regard to this situation, we only wish to describe how a field ontology may be capable of providing a sensible ontological framework for EFTs. 

Given that an EFT is valid only down to some distance scale, it should be able to provide an answer to the question, ``what does the world look like, at this resolution?'' without committing to the details at higher resolutions. With a field ontology, one readily has an answer, in principle: one can imagine observing a ``large-scale'' lump structure, but zooming in on any small patch of the lump will yield non-physical structures, structures that could equally-well have been characterized differently. This is manifested on the lattice toward a continuum limit, where the spacing $a$ is made much smaller than the inverse of $m_R$, the renormalized mass, and the fluctuations in the field on scales of a few lattice spacings are not regarded as physical. We have seen that Nelsonian QFT provides a way to actualize this sort of ontology. Nelson's framework has the further property that it treats the evolution of the field in an effective way, by characterizing the interaction of the beable with a background field through osmotic and noise terms in the guiding law, in much the same way that Brownian motion is an effective way to treat the motion of a particle in a liquid without knowing the detailed motions of all the particles that collide with it. However, because it is commonly thought that EFTs can form a \emph{tower}, one would have to argue, in order to demonstrate a complete assimilation of Nelson into EFT, that the effective treatment of the background field on one level arises from an appropriately coarse-grained description of a field theory on a deeper level. And this remains to be done.

\paragraph{Acknowledgements.} I would like to thank Maaneli Derakhshani, Ward Struyve, and Charles Sebens for discussions and feedback. This work has also benefited from comments received after presenting the material at the 2023 European Conference on Foundations of Physics in Bristol, UK, as well as the online Laws of Nature Series \cite{laws-of-nature-channel}.

This work was supported in part by U.S. DOE Grant No. DE-FG02-95ER40907.

\bibliographystyle{ieeetr}
\bibliography{Refs.bib}

\end{document}